%% file: main.tex
\documentclass[sigconf,nonacm,xcolor,dvipsnames,table]{acmart}

\newcommand\vldbdoi{XX.XX/XXX.XX}

 \usepackage{relsize}
\usepackage{subcaption}
\usepackage{tabularx}
\usepackage{afterpage}
\usepackage{soul}
\usepackage{balance}
\usepackage{footmisc}
\usepackage[skip=0.5pt]{caption}
\usepackage[export]{adjustbox}
\usepackage[linesnumbered,ruled,noend]{algorithm2e}
\usepackage{listings}
\usepackage{standalone}
\usepackage{enumitem}
\usepackage[noabbrev,nameinlink,capitalise]{cleveref}
\usepackage{multirow}

\newcommand{\ag}[1]{{\color{blue}AG: #1}}

\newcommand{\paratitle}[1]{\noindent{\bf #1}.}

\newcommand{\sys}{{\textsc{fedex}}}
\newcommand{\system}{\sys}
\newcommand{\sysopt}{{\textsc{fedex-Sampling}}}
\newcommand{\seedb}{{\textsc{SeeDB}}}
\newcommand{\io}{{\textsc{IO}}}
\newcommand{\rath}{{\textsc{Rath}}}
\newcommand{\expert}{{\textsc{Expert}}}

\newcommand{\df}{d}

\newcommand{\inputdf}{d_{in}}
\newcommand{\inputdfs}{D_{in}}
\newcommand{\outputdf}{d_{out}}

\newcommand{\attr}{A}

\newcommand{\acolumn}{\df[A]}
\newcommand{\acolumnin}{\inputdf[A]}
\newcommand{\acolumnout}{\outputdf[A]}

\newcommand{\opr}{q}

\newcommand{\gopr}{q_g}

\newcommand{\eat}[1]{}

\AtBeginDocument{%
  \providecommand\BibTeX{{%
    \normalfont B\kern-0.5em{\scshape i\kern-0.25em b}\kern-0.8em\TeX}}}

\renewcommand\footnotetextcopyrightpermission[1]{} 

\newcommand{\reva}[1]{{\leavevmode\color{black}{#1}}}
\newcommand{\revb}[1]{{\leavevmode\color{black}{#1}}}
\newcommand{\revc}[1]{{\leavevmode\color{black}{#1}}}
\newcommand{\shep}[1]{{\leavevmode\color{black}{#1}}}
\newcommand{\common}[1]{{\leavevmode\color{black}{#1}}}

\begin{document}
\fancyhead{} 

\title{\sys: An Explainability Framework for Data Exploration Steps}

\setcounter{figure}{0}






\author{Daniel Deutch$^\dagger$, Amir Gilad$^\ddagger$, Tova Milo$^\dagger$, Amit Mualem$^\dagger$, Amit Somech$^\mathsection$}
\email{danielde@post.tau.ac.il, agilad@cs.duke.edu, milo@cs.tau.ac.il, amitmualem@mail.tau.ac.il, somecha@cs.biu.ac.il}
\affiliation{$^\dagger$Tel Aviv University, $^\ddagger$Duke University, $^\mathsection$Bar-Ilan University}

\begin{abstract}
    When exploring a new dataset, Data Scientists often apply analysis queries, look for insights in the resulting dataframe, and repeat to apply further queries. We propose in this paper a novel solution that assists data scientists in this laborious process. In a nutshell, our solution pinpoints the most interesting (sets of) rows in each obtained dataframe. Uniquely, our definition of interest is based on the contribution of each row to the interestingness of different columns of the entire dataframe, which, in turn, is defined using standard measures such as diversity and exceptionality. Intuitively, interesting rows are ones that explain why (some column of) the analysis query result is interesting as a whole. Rows are correlated in their contribution and so the interesting score for a set of rows may not be directly computed based on that of individual rows. We address the resulting computational challenge by restricting attention to semantically-related sets, based on multiple notions of semantic relatedness; these sets serve as more informative explanations. 
    Our experimental study across multiple real-world datasets shows the usefulness of our system in various sc\textbf{}enarios. 
\end{abstract}

\maketitle
\setcounter{page}{1}



\input{introduction_revision}
\input{related}


\input{solution_new}
\input{experiments} 
\input{conclusions}

\begin{acks}
This research has been partially funded by the European Research Council (ERC) under the European Union’s Horizon
2020 research and innovation programme (grant agreement No. 804302), the Israel Science Foundation, BSF - the Binational US-Israel Science foundation and the Tel Aviv University Data Science center, and the NSF awards IIS-1703431, IIS-1552538, IIS-2008107. 
\end{acks}

\clearpage

\balance
\bibliographystyle{ACM-Reference-Format}
 \small{
 \bibliography{bibtex}
 }  
 
\clearpage 
\input{appendix}
\end{document}

%% file: introduction_revision.tex
\section{Introduction}\label{sec:intro}


Exploratory Data Analysis (EDA) is an essential process performed by data scientists and analysts in order to up-close examine a new dataset, better understand its nature and characteristics, and extract preliminary insights from it. 
EDA is typically performed in \textit{scientific notebooks}~\cite{kery2018story} using a dataframe library~\cite{reback2020pandas} which allows users to interactively transform and analyze datasets via programmatic query operations (such as filter, aggregation, join, pivot, etc). 
In each exploratory step, the analyst runs a query over the previously obtained dataframe, examines the resulting dataframe to derive intermediate insights, and decides on the next exploratory step. 


EDA is a challenging process, where arguably the most significant challenge lies in scanning the (possibly many) query results obtained in each step to identify interesting patterns and trends. These discoveries are crucial both for gathering insights and conceiving the next exploratory steps.  
To illustrate the difficulty faced by analysts in interpreting the results of exploratory steps, consider the following example EDA scenario.

\begin{figure*}[t!]
{\footnotesize
\begin{subfigure}[b]{0.7\linewidth}
\begin{center}
\ttfamily
\begin{tabular}{|l|l|c|c|c|c|c|c|c|c|}
    \hline
    \rowcolor{gray!40}
    \textbf{name} & \textbf{main\_artist} & \textbf{year} & \textbf{decade} & $\ldots$ & \textbf{danceability} &  \textbf{loudness} & \textbf{popularity} \\
    \hline
    Maybe I'm Amazed... & Paul McCartney & 1970 & 1970 & $\ldots$ & 0.471 & -10.407 & 66 \\
    Life on Mars?... & David Bowie & 1971 & 1970 & $\ldots$ & 0.442 & -14.635 & 71 \\
    Time in a bottle & Jim Croce & 1972 & 1970 & $\ldots$ & 0.544 & -11.952 & 67 \\ 
    Desperado - 2013... & Eagles & 1973 & 1970 & $\ldots$ & 0.228 & -12.749 & 67 \\ 
    $\ldots$ & $\ldots$ & $\ldots$ & $\ldots$ & $\ldots$ & $\ldots$ & $\ldots$ & $\ldots$ \\ \hline
\end{tabular}
\end{center}
\caption{Filter results}
\label{fig:spotify_filter}
\end{subfigure}%
\begin{subfigure}[b]{0.3\linewidth}
\begin{center}
\ttfamily
\begin{tabular}{|c|c|c|l|l|}
    \hline
    \rowcolor{gray!40}
    \textbf{year} & \textbf{loudness} & \textbf{danceability} \\
    \hline
    1991 & -11.076072 & 0.555551 \\
    2014 & -7.826855 & 0.586153 \\
    1992 & -10.694651 & 0.555135 \\
    2013 & -8.238654 & 0.593881 \\
    $\ldots$ & $\ldots$ & $\ldots$ \\ \hline
\end{tabular}
\end{center}
\caption{Group-by results}
\label{fig:spotify_groupby}
\end{subfigure}
}
\vspace{-3mm}
\caption{\linespread{1}\selectfont{} \textbf{Result samples of filter and group-by operations over the Spotify ``Song Popularity Dataset''.} \normalfont{Figure~\ref{fig:spotify_filter} shows partial results of a filter operation that leaves in only songs with `popularity' scores above 65. Figure~\ref{fig:spotify_groupby} depicts partial results of a group-by operation, showing for each year -- since 1990 -- the mean `loudness' and `danceability' values.}}
\label{fig:spotify}
\end{figure*}

\begin{figure*}[t]
\centering
\begin{subfigure}[b]{0.49\textwidth}
\fbox{\includegraphics[width=0.92\linewidth]{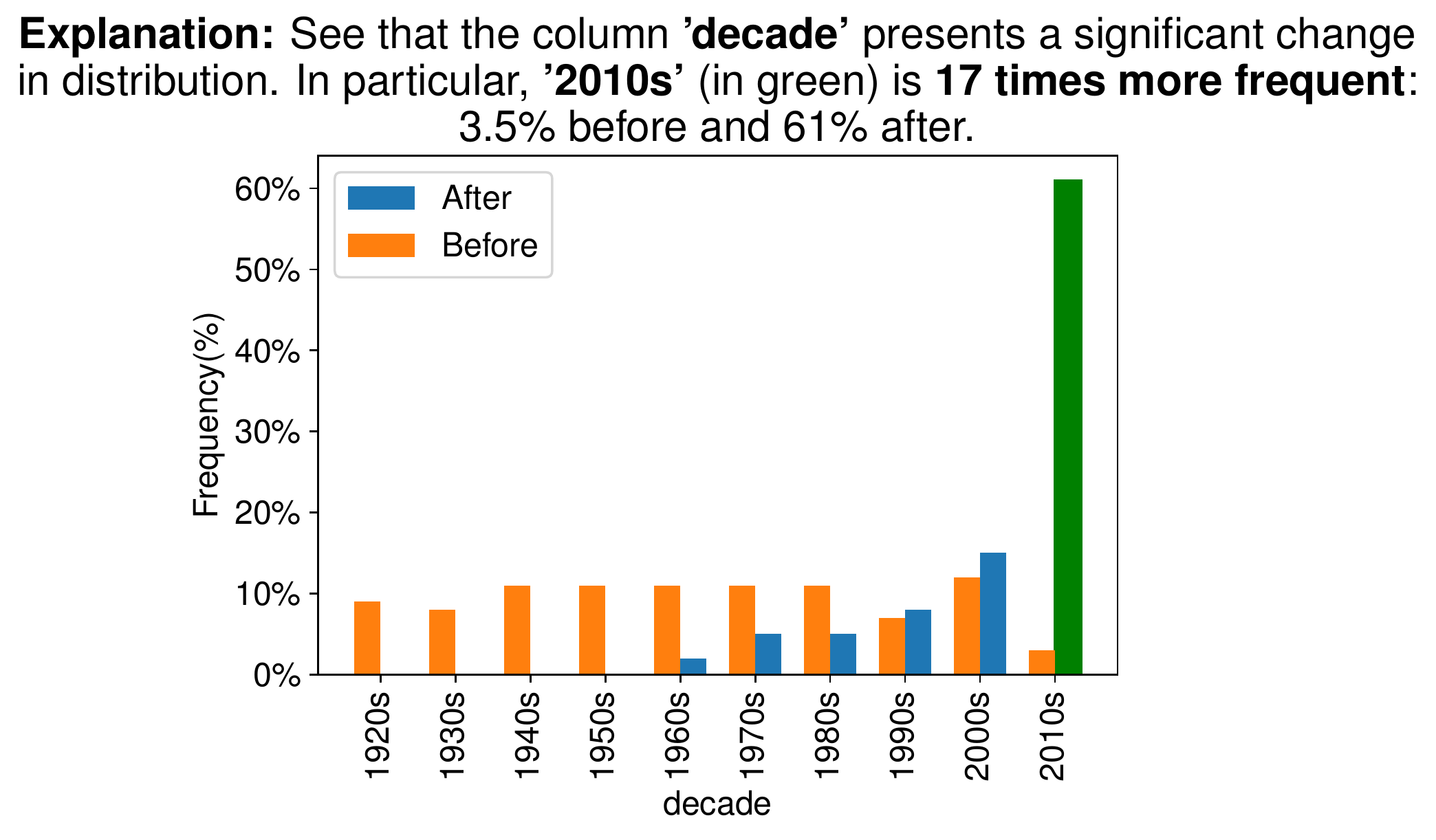}}
\vspace{-2mm}
    \caption{Explanation for the filter step (Figure~\ref{fig:spotify_filter})}
    \label{fig:explained_filter}
    \end{subfigure}
    \hspace*{\fill}
\begin{subfigure}[b]{0.49\textwidth}
\vspace{-52mm}
    \fbox{\includegraphics[width=0.97\linewidth]{{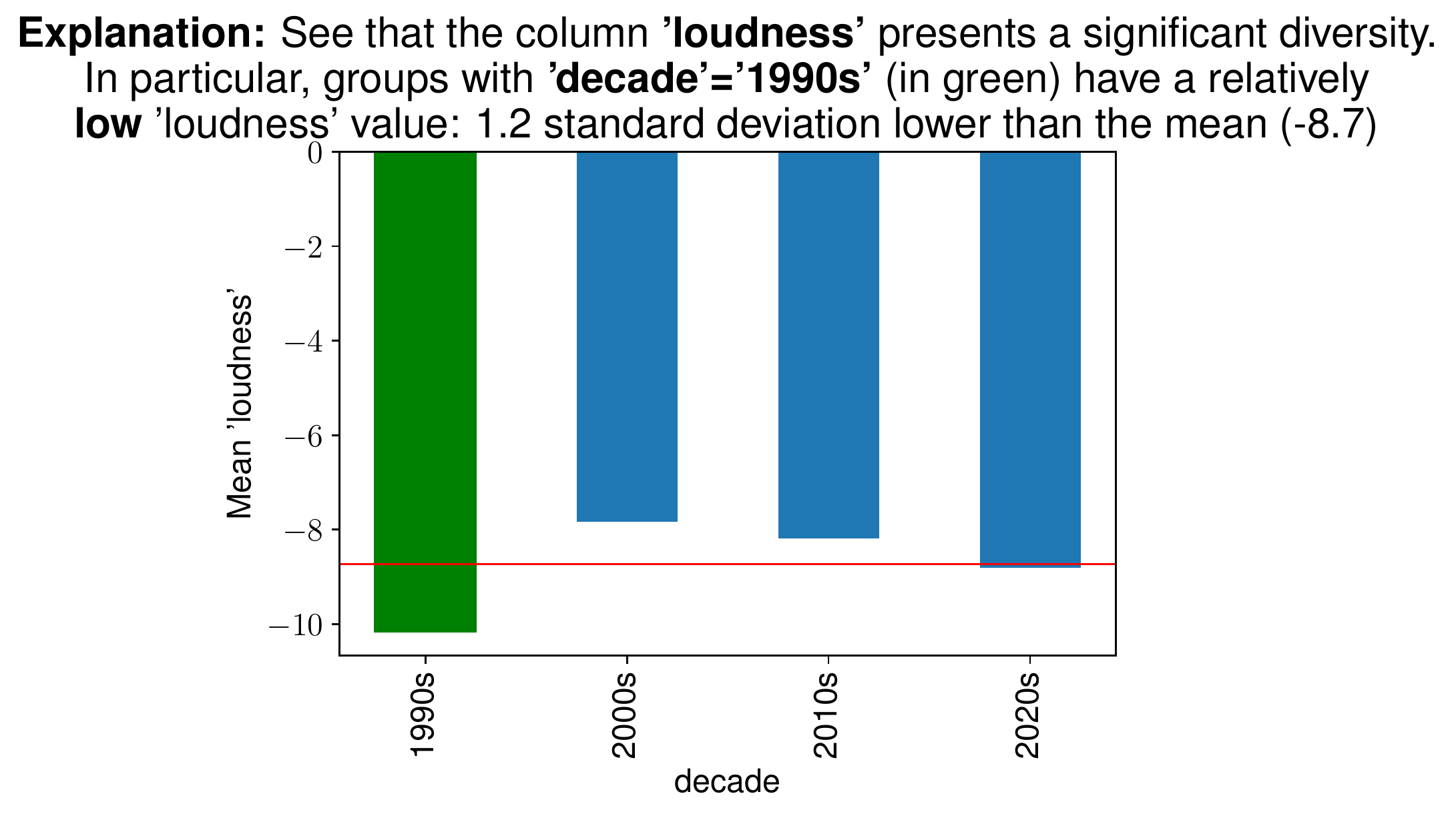}}\vspace{8mm}}
    \vspace{-2mm}
     \caption{Explanation for the group-by step (Figure~\ref{fig:spotify_groupby})}
     \label{fig:explained_groupby}
    \end{subfigure}
    \caption{  \textbf{Multi-modal explanations generated by \sys{}.} \normalfont{Figure~\ref{fig:explained_filter} explains the filter results in Figure~\ref{fig:spotify_filter}. Our framework has identified that the highest contribution to the interestingness score is due to songs made in the 2010s --  they form 61\% of the popular songs, compared to only 3.5\% in the entire dataset, leading to the insight that \underline{\textit{new songs tend to be more popular than older songs}}}. Figure~\ref{fig:explained_groupby} explains group-by results in Figure~\ref{fig:spotify_groupby}. Our explanation points out that \underline{\textit{songs made in the 90s tend to be less loud than later songs}}.} 
    \label{fig:explained_new}
\end{figure*}

\begin{example}\label{ex:spotify-eda}
Data scientist Clarice works on a dataset of song information, published by Spotify\footnote{\url{https://www.kaggle.com/c/bfh-spotify-challenge/overview}}.

The dataset (see samples in Figure~\ref{fig:spotify}) contains a \textit{popularity score} for each song, alongside 37 descriptive features  (e.g., artist name, year, decade) as well as audio-analysis features such as \textit{loudness}, \textit{danceability}, and others. 

\revb{Clarice is first interested in the question ``what makes songs popular?''. She, therefore, composes a filter operation showing songs that have a \textit{high} popularity score ($>65$).}
Applying this operation yields a dataframe with 7000 rows and 37 columns (see Figure~\ref{fig:spotify_filter} for a small sample). Clarice now needs to examine these results in order to understand what is interesting about them. One such notion of interestingness may be captured by the question ``In what way are these popular songs different than the rest of the songs in the dataset?" Clarice therefore needs to sift through the rows, apply additional data transformations, and data visualizations.

\revb{Clarice then decides to focus her analysis only on recent songs (released after 1990) and investigate their characteristics. 
To do so, she first filters the original dataframe to output songs released after 1990,
and then performs a group-by operation to view the mean \textit{loudness} and \textit{danceability} values for each year.}

A sample of the result of this group-by step appears in Figure~\ref{fig:spotify_groupby}. This step results in a much smaller dataframe than before, 
including about 30 rows and 3 columns. However, interesting patterns and trends are still not clearly visible -- are there any particular years with more `quiet' songs than the other? is there a trend where newer songs are more `loud' and `danceable'? Answering these questions still requires a substantial effort.  
\end{example}




In this paper, we present \sys{} (Ef{\bf f}ici{\bf e}nt {\bf D}ata Exploration {\bf Ex}planations), an EDA explanation framework that assists users in analyzing and understanding the results of their exploration steps. \sys{} explains exploratory steps using a twofold process: (1) as an analyst would do, \sys{} inspects the resulted dataframe and discover interesting aspects of it. For example, does it show outliers? Or perhaps a highly diverse set of values in one of the columns?; (2) since naturally, not all underlying tuples equally contribute to the interesting pattern discovered, \sys{} detects which data subsets \textit{cause} the resulted dataframe to be interesting. Therefore, a good \textit{explanation} for an exploratory step is a set-of-rows from the dataframe which \textit{significantly contributes} to a \textit{highly interesting} aspect of the results dataframe.  
Importantly, \sys{} tailors the explanation to the {\em context} of the generated dataframe, i.e., the EDA operation or query and the source dataframe.
\sys{} then presents this set-of-rows in a hybrid format via the notebook interface as coherent, easy-to-read \textit{captioned visualizations}. This allows users to quickly understand and derive immediate insights from each exploratory step they make. 

\begin{example}[\sys\ Explanations]\label{ex:explanations}
Figure~\ref{fig:explained_new} shows the explanations produced by \sys{}, for the two exploratory steps on the Spotify dataset in Example \ref{ex:spotify-eda} (see Figure~\ref{fig:spotify}). 
For the filter step, which focuses only on songs with a popularity score greater than 65, \sys{} detects that in the `decade' column there exists a high, interesting \textit{deviation} from the input dataframe (namely, the dataframe Clarice applied the filter on). It then detects that most of the deviation is due to \textit{songs from the \textit2010s}. Using a side-by-side bar plot, \sys{} highlights the difference in the number of songs from the 2010s decade before and after the filter, showing that \textit{newer songs from the last decade tend to be more popular than older songs}. 
As for the group-by step, showing the mean `loudness' and `danceability' values for each release year, \sys{} identifies an interesting pattern in the column `loudness', whose values greatly differ from one another. It then detects that the large diversity in values occurs mainly due to years in `1990s', highlighting that \textit{the songs released in the 1990s are significantly less loud than songs made in later years}. 
\end{example}

Multiple approaches exist for assisting data exploration,
such as recommending the next exploratory steps~\cite{Somech2018REACTKDD,seleznova2020guided},
query auto-completion~\cite{khoussainova2010snipsuggest,le2019explique}, visualization~\cite{VartakRMPP15,lee2021lux,wongsuphasawat2016voyager} and even insights~\cite{tang2017extracting,asudeh2020detecting} recommendations. These works reduce the manual effort of composing queries, yet to the best of our knowledge, \sys\ is the first system to automatically generate explanations for output dataframes {\em in the context of exploratory steps.}

Our main contributions can be summarized as follows. 

\begin{itemize}[leftmargin=0.2in, topsep=1pt]
    \item 
    
    We devise \sys, a novel explainability framework for data exploration, using a hybrid, twofold approach: (1) \sys\ uses measures of interestingness \cite{sarawagi1998discovery,hilderman2013knowledge} to detect interesting columns in the resulted output dataframe. (2) \sys\ then applies a notion of \textit{contribution of a set of rows}, inspired by previous work \cite{0002M13}, to identify meaningful subsets of the input dataframe that significantly contribute to the interestingness score of one of the columns in the output dataframes.  We define an explanation candidate as a tuple $(R,A)$, where $R$ is a set of rows in the input dataframe that contribute to the interestingness of a column $A$ in the step's output dataframe. 
    
    \item 
     A naive computation of explanation candidates would imply exponential complexity as we would have to traverse all possible pairs of columns and row subsets. Therefore, \sys\ employs the following efficient process for finding the most promising explanations in terms of both interestingness and contribution, which are also semantically meaningful:
    First, \sys\ uses a two-step greedy approach to calculate the most promising explanation candidates: (1) it identifies the interesting columns, utilizing a sampling-based calculation of interestingness, which significantly accelerates the computation while maintaining the quality of the results. On the resulted interesting columns, it then (2) computes the contribution of only \textit{selected} sets of rows (rather than all of them) by partitioning the rows into \textit{semantically meaningful sets-of-rows}. This is done by several automatic partitioning methods 
    which dramatically reduce the number of candidate and keep only the meaningful ones.
    Finally, to return the most useful explanations,  \sys\ uses the skyline operator~\cite{borzsony2001skyline} over the compact set of semantically-meaningful explanation candidates, to retrieve only the \textit{dominating} ones, in terms of contribution and interestingness scores. Each skyline explanation is presented 
    to the user via a dedicated captioned visualization.
    


    
    \item 
    We have implemented a prototype version of \sys{} (available in \cite{repo}), performed comprehensive user studies, and simulated experiments. The user studies include a detailed comparison of our approach to four baselines, based on previous work as well as on explanations generated manually by experts. In our simulated experiments, we examine the scalability of our approach and further considered the usefulness of our optimization in aiding scalable performance while maintaining high accuracy. 
    The results indicate that our approach is 1.7 times more helpful than commonly used baselines on average, allows users to get approximately 4 more insights on average than unassisted EDA, and runs at interactive speed for large data. 
    
\end{itemize}

%% file: related.tex
\section{Related Work}
\label{sec:related} 
As mentioned above, multiple lines of work have studied solutions to assist data scientists in data exploration, 
focusing mainly on the aspect of formulating and composing exploratory operations. For example, 
systems for query suggestions~\cite{Somech2018REACTKDD,seleznova2020guided}, simplified EDA interfaces for non-programmers~\cite{srinivasan2018augmenting,singhdbexplorer,bao2015exploratory,bespinyowong2016exrank}, query formulation assistance ~\cite{khoussainova2010snipsuggest,le2019explique}, \reva{and EDA guidance tools~\cite{sarawagi2001user,joglekar2017interactive}}. 
We focus here on the challenge of {\em explaining the results of the exploratory steps and assisting users in quickly understanding them.} In this context, we survey three lines of related work: (1) interestingness assessment in data exploration, (2) query explanation frameworks, and (3) automatic visualizations and insights discovery.

\paratitle{Modeling and predicting interest in exploratory sessions} Modeling interestingness in EDA, i.e., assessing whether a resulted view of an EDA operation is interesting or not, is a known challenge: previous work showed that interestingness in EDA is multi-faceted~\cite{mcgarry2005survey,geng2006interestingness}, often subjective~\cite{de2013subjective} and dynamically changing, even in the same exploratory session~\cite{milo2019predicting}. 

Consequently, several works attempt to model or predict user interests in EDA sessions: works such as \cite{milo2019predicting,luo2018deepeye} propose predicting the most suitable notion of interestingness by mining logs from previous EDA sessions. The system described in \cite{milo2019predicting} is used to dynamically select, before the user employs another EDA operation, the most suitable among a set of \textit{existing} interestingness measures, whereas in~\cite{luo2018deepeye} a \textit{learning-to-rank} model is used to construct an ad hoc interestingness notion. 
Another line of work proposes modeling user interests by collecting live feedback: systems such as~\cite{qian2015learning,dimitriadou2016aide} ask the user, e.g., to annotate presented tuples as ``interesting'' or ``non-interesting'', while \textit{explore-by-example} systems~\cite{deutch2016qplain,sellam2016cluster} ask the users to provide examples for interesting tuples in advance.

\revb{
 While \sys\ employs measures of interestingness, the end goal is fundamentally different. 
Rather than using these measures to recommend interesting operations or views as in previous work, \system{} uses the measures to evaluate the users' operations and generates explanations based on sets-of-rows that impact the interestingness. This allows \sys{} to identify \textit{why} a user's operation is interesting.}  Furthermore, in Section~\ref{sec:user-study} we show that explanations generated by directly applying  interestingness measures are generally less useful compared to those generated by \sys{}.



\paratitle{Explaining query results}
Explaining the results of SQL queries for debugging purposes has been extensively researched within the database community. 
Prominently, such research works utilize data provenance, as well causality-stemmed notions such as intervention and influence, in order to identify tuples whose existence or absence affects the result of the inspected query. For example,~\cite{GKT-pods07,why} explain the \textit{existence} of particular tuples in the query result, whereas works e.g. ~\cite{chapman2009not,bidoit2014query,bidoit2015efficient,ten2015high} explain the \textit{absence} of particular tuples. Explaining aggregation results has been suggested in~\cite{amsterdamer2011provenance}, and outliers in  ~\cite{0002M13,RoyS14,shafieinejad2021pcor}. In addition, works e.g., ~\cite{miao2019going,li2021putting,DeutchFG17} propose augmenting provenance with additional sources of information such as counter-balancing patterns~\cite{miao2019going}, related tables~\cite{li2021putting}, and natural language questions~\cite{DeutchFG17}. 
\reva{Last, data debugging tools also utilize notions of causality and intervention, e.g., to detect erroneous tuples that significantly impact an aggregation result~\cite{barowy2014checkcell}, and to detect faults in predictive models that may be caused by non-conforming data subsets~\cite{fariha2020extune}. }

 \reva{The main difference from \system{} is that these works explain which input tuples affect the \textit{direct output} of a query (i.e., why certain tuples \shep{appear/do not appear} in the results), whereas the goal of our work is
 focused on explaining \textit{why} an EDA operation is \textit{interesting}, by detecting sets-of-rows that positively impact the interestingness of users' exploratory steps.}

\paratitle{Automatic visualization generation~\& insights discovery}
In recent years, a plethora of visualization generation systems have been devised~\cite{VartakRMPP15,qin2020making,wongsuphasawat2016voyager,zegarra2020visual,luo2018deepeye,behar2020optimal_cikm,behar2020optimale_edbt}, taking a dataset as input and generating useful data visualizations (e.g., histograms, line plots, heatmaps). Many such systems utilize a notion of \textit{utility} or \textit{importance}, and scan the input dataset to show the top-k visualizations obtaining the highest score~\cite{VartakRMPP15,qin2020making,wongsuphasawat2016voyager}.


Using roughly similar methods, previous works propose solutions for facts and insights discovery. Given an input dataset, these solutions can automatically detect several types of useful insights, such as \textit{outliers}~\cite{ilyas2004cords,tang2017extracting}, \textit{rare categories}~\cite{huang2019leri}, \textit{conditional correlations}~\cite{mirylenka2013finding,mirylenka2015conditional}, and \textit{trend-lines}~\cite{asudeh2020detecting,tang2017extracting,ding2019quickinsights}. Other solutions propose insights discovery for \textit{specific} use-cases such as \textit{user rating data}~\cite{thirumuruganathan2012maprat}, \textit{smart meters data}~\cite{liu2015smas}, \textit{OLAP}~\cite{agarwal2007efficient} and \textit{data fusion}~\cite{dong2013compact}.

A step further, systems like~\cite{cui2019text,wang2019datashot} generate random interesting facts (in a similar manner to, e.g.,~\cite{tang2017extracting,ding2019quickinsights}), then organize them in a comprehensive medium such as fact sheets~\cite{wang2019datashot,shi2020calliope} and infographics~\cite{cui2019text} or use them as guidance in interactive exploration~\cite{srinivasan2018augmenting}.




%

Similarly, \sys{} also assists users in discovering insights from the data. However, rather than generating random, arbitrary facts from the dataset (as e.g.,~\cite{VartakRMPP15,qin2020making,wongsuphasawat2016voyager,wang2019datashot,tang2017extracting}), \sys{} generates insights that explain each exploratory step made by the user. The methods used in \sys{} are also different than previous work, and ours is the only work, to our knowledge, that derives insights by interweaving interestingness and causality assessment.
In our experimental evaluation, we compare the explanations of \sys{} to visualizations and insights generated by ~\cite{VartakRMPP15} and ~\cite{wang2019datashot}, and show that \sys{} is significantly more effective in explaining  exploratory steps. 






%% file: solution_new.tex
\section{Explanation Framework}\label{section:algorithms}
We begin with a brief overview of our data model for notebook-based exploration, then detail each of the components in \sys{}.

\subsection{Model for Notebook-based EDA}
\label{sec:model}
As mentioned above, the notebook interface allows users to explore datasets interactively using a dataframe module~\cite{reback2020pandas}. The user typically loads a dataset
as a dataframe from an external source (e.g., a CSV file, spreadsheet, or an SQL query result from a database server), then interacts with them by employing programmatic exploratory operations, such as filter, group-by, aggregate, etc. The results of EDA operations are returned to the user also as dataframes; these dataframes may be viewed and/or used as input for subsequent exploration steps. 

\paratitle{Dataframe}
Formally, a dataframe $\df$ is equivalent to a single relational table or view. It comprises of a multiset of rows over a schema $\mathcal{A}(\df)$. 
Each row $r \in \df$ represents either a tuple or a group (in case $\df$ is resulted from a group-by operation). Correspondingly, an attribute $\attr \in \mathcal{A}(\df)$ specifies either a table attribute, or an aggregated column. 
Furthermore, we refer to a dataframe \textit{column} by $\acolumn$, i.e., the multiset of all rows' values associated with the attribute $\attr$ in dataframe $\df$. 

\paratitle{EDA operations} 
\reva{ We consider popular, commonly used (see \cite{YanH20}) of filter, join, group-by, and union.
} Additional, advanced EDA and OLAP operations such as \textit{pivot}, \textit{diff}, and \textit{roll-up} can be supported by a simple extension of our model.
\revb{We further denote a full exploratory step 
by $Q=(\inputdfs,q,\outputdf)$, i.e., applying operation $q$ on an input dataframe(s) $\inputdf (\inputdfs)$, resulting in an output dataframe $\outputdf$. Multiple input dataframes ($|\inputdfs| > 1$) may be used when the operation is a join or union.} 

The following example demonstrates the workflow of a notebook-based EDA process.

\begin{example}
Reconsider the Spotify songs dataset, described in Example \ref{ex:spotify-eda}.
After a user loads it into a dataframe $\df_0$, she performs the first exploratory step
by employing a \textit{filter} operation $q_1$ equivalent to the SQL query \textit{'select * from $\df_0$ where popularity>65'} 
This results in a new dataframe $\outputdf= q_1(\df_0)$, shown in Figure~\ref{fig:spotify_filter}, that includes the subset of rows that comply to the criterion of ``popularity$>$65''. 


The group-by operation from our running example, specified by $q_2$, is equivalent to the query \textit{`select AVG(loudness), AVG(danceability) from $\df_0$ where year>=1990 group by year'}


The resulting dataframe (depicted in Figure~\ref{fig:spotify_groupby}) shows the mean values for \textit{loudness} and \textit{danceability} per year, starting from 1990.


\end{example}



\subsection{Interestingness of Exploration Steps}
\label{sec:interestingness}



Measuring interestingness of data analysis operations has been intensively discussed in previous work (see \cite{geng2006interestingness,hilderman2013knowledge} for surveys). These works define numerous abstract facets of interestingness, such as novelty, surprisingness, exceptionality, and diversity.
Often such notions are implemented differently, depending on the data and task~\cite{VartakRMPP15,qin2020making,wongsuphasawat2016voyager}.
\reva{\sys{} can take an existing or a custom, user-defined notion of interestingness (See Section~\ref{sec:generalization}). As default measures, \sys{} utilizes two interestingness functions corresponding to the notion of \textit{exceptionality} and {\em diversity}. Similar implementations of such functions were proven useful in the context of EDA tasks~\cite{van2010maximal,VartakRMPP15,bar2020automatically,milo2019predicting}.
}
\reva{These two measures are applicable, as described below, for all the EDA operations supported by \sys\ (filter, join, group-by, and union).}



Since applying $q$ does not equally affect all columns in the output dataframe $\outputdf$, following~\cite{Somech2018REACTKDD,milo2019predicting}, we assess the interestingness of the step $Q=(\inputdfs,q,\outputdf)$ separately for each column in the output dataframe $\outputdf$. 
We denote by $I_A(Q)$ the \textit{interestingness score} of step $Q$, by an interestingness function $I$, w.r.t. attribute $A$. 

\paratitle{Exceptionality (Filter/Join/\reva{Union})}
Inspired by ~\cite{sarawagi1998discovery,van2010maximal,VartakRMPP15}, we introduce an exceptionally measure that is particularly suitable for filter and join steps, as follows.
For the case of a filter operation $q_f$, 
this measure intuitively deems the filter result interesting if it produces an output dataframe which significantly \textit{deviates} from the input dataframe. Namely, the filter operation caused a significant change in the distribution of the values of a given column. 
We, therefore, quantify the deviation of the step $Q=(\inputdf,q_f,\outputdf)$ w.r.t Attribute $A \in \mathcal{A}(\outputdf)$ by measuring the differences 
between the values of $\acolumnin$ and $\acolumnout$, which are the columns associated with attribute $\attr$ in dataframes $\inputdf$ and $\outputdf$ (resp).
Concretely, we use the two-sample Kolmogorov–Smirnov (KS) test \cite{ross2004introduction},
a well-known statistical test used to assess whether two samples originate from the same probability distribution.
First, we define the column probability distribution $Pr(\acolumn)$ based on the relative frequency of its values (i.e., for each value $v \in \acolumn$, $Pr(v)$ is the probability to choose $v$ uniformly at random). 
We then calculate the exceptionality-based interestingness score:
\begin{equation}
\label{eq:ks}
    I_A(\inputdf,q_f,\outputdf) := KS(Pr(\acolumnin),Pr(\acolumnout))
\end{equation}

\revb{The same measure is also suitable for quantifying the interestingness of the join \reva{and union} operations since, oftentimes, the results of a join \reva{(or union)} contain a different number of tuples than in the original input dataframes (e.g., if not all tuples in the input dataframes match on the join condition). 

For a join operation $\opr_j$ and a column $A$ in the output dataframe $\outputdf$, 
the interestingness is measured by
$I_A(\inputdf',\opr_j,\outputdf)$,
where $\inputdf'$ is the corresponding input dataframe in $\inputdfs$, s.t. $A \in \mathcal{A}(\inputdf')$ and $I_A$ is calculated as in Equation~\ref{eq:ks}. 
}

\reva{For a union operation $\opr_U$, 
the interestingness of a column $A$ is defined by
 $\max_{\inputdf \in \inputdfs} I_A(\inputdf,\opr_U,\outputdf)$,
namely the maximal KS difference of the output column $\outputdf[A]$ and each of the input dataframes to be unionized.}





\paratitle{Diversity (group-by)}
We have also implemented an interestingness function that captures \textit{diversity}~\cite{hilderman2013knowledge,bedeian2000use}. 
It is particularly suitable for group-by operations, since, intuitively, a group-by step that yields a dataframe with a highly diverse set of aggregated values, implies a large difference between the groups. 
Therefore, our interestingness function quantifies the diversity of a column in the output dataframe of a group-by operation $q_g$ by calculating the \revb{coefficient of variation (CV)~\cite{bedeian2000use} of the aggregated values: 
\begin{equation}
\label{eq:var}
I_A(\inputdf,\gopr,\outputdf) := CV(\acolumnout) = \frac{1}{\bar{a}} \cdot \sqrt{\frac{\sum(a_i - \bar{a})^2}{n-1}}
\end{equation} 
}

Where the sum in the numerator is over $a_1,a_2,\dots,a_n$ which are the aggregated values in column $A$ after employing $q_g$, $\bar{a}$ is mean value, and $n$ is the number of groups. 
The following example demonstrates the usage of the exceptionality and diversity measures as defined above, to evaluate the interestingness of the filter and group-by operations shown in Figure~\ref{fig:spotify}.
 

\begin{example}
\label{example:interestingness}
Consider again the filter and group-by results depicted in Figure~\ref{fig:spotify}. For the group-by results (Figure~\ref{fig:spotify_groupby}),
it is clearly visible that the values in column `loudness' are more diverse than that of  `danceability', which shows value distribution tightly around $0.55$. Indeed, the diversity-based score of the column `loudness' is \revb{$0.13$}, whereas the score of the column `danceability' is \revb{$0.04$}. Hence, our framework will focus on the former column to explain the interestingness of the view (See Figure~\ref{fig:explained_groupby}).
 As for the filter results, while this is not visible in the sample in Figure~\ref{fig:spotify_filter}, the highest deviation was measured for the column `decade', obtaining an interestingness score of $0.56$, followed by the columns `year' and `loudness', for which the scores are $0.54$ and $0.41$, respectively.

 \end{example}

%

\begin{table}[!t]
\caption{Notations Summary} 
\centering
{\small
\begin{tabular}{|p{0.13\textwidth}| |p{0.3\textwidth}|p{0.1\textwidth}}
\hline
\textbf{Notation} & \textbf{Description} \\ \hline
 $d$, $\mathcal{A}[\df]$ & Dataframe, Dataframe Schema \\ \hline 
$\acolumn,~A \in \mathcal{A}[\df] $ & Dataframe column   \\ \hline
$q$ & EDA operation specifications   \\ \hline
\revb{$\inputdf (\inputdfs), \outputdf$} & \revb{Input Dataframe(s), Output Dataframe}  \\ \hline
$Q=(\inputdfs,q,\outputdf)$ & Exploratory step \\ \hline
$I_A(Q)$ & Interestingness of step $Q$ w.r.t attribute $A$ \\ \hline
$R \subseteq \df$ & Set-of-Rows in a dataframe $\df$  \\ \hline
\multirow{2}{*}{$C(R,A,Q)$} & Contribution of a set-of-rows $R$ to interestingness of a column $A$ in step $Q$  \\ \hline
$\mathcal{R} = \{R_1,R_2,\dots,\hat{R}\}$ & Row partition of a dataframe to sets-of-rows \\ \hline
\multirow{2}{*}{$\Bar{C}(R,A)$} & Standardized contribution of a set-of-rows $R\in \mathcal{R}$, compared to the rest of the row-sets in $\mathcal{R}$ \\ \hline
\multirow{2}{*}{$(R,A) \in EC$} & The set of explanation-candidates, each is a pair of set-of-rows and a single column in $\outputdf$ \\ \hline

\multirow{2}{*}{$E \in EX$} & A dominating explanation in terms of contribution and interestingness  \\ 
 \hline
\end{tabular}
\label{table:algorithms}
\label{table:notations}
}
\end{table}

\subsection{Contribution of Sets of Rows}\label{sec:contribution}



Given an exploratory step $Q=(\inputdfs,q,\outputdf)$, we further aim to focus the explanation by quantifying the \textit{contribution} of a set-of-rows $R \subset \inputdfs$ to the interestingness of a column $A$, $I_A(Q)$.
Hence, we define a contribution function, as follows. 
Our definition of contribution draws on the notion of intervention, first defined in the context of causality \cite{pearl2009causality} to measure the change in the outcome when the input changes. This notion was adopted by the database community for measuring the change in query results when some of the rows are absent \cite{meliou2010causality,0002M13,RoyS14}. 
In particular, our notion of contribution function is inspired by \cite{0002M13} and defined as follows.

\begin{definition}[Contribution of a set-of-rows] For a given step $Q=(\inputdfs,q,\outputdf)$, the amount that a set-of-rows $R \subset \inputdfs$ \textit{contributes} to the interestingness  of column $A \in \mathcal{A}(\outputdf)$ is calculated by: 
\[C(R, A, Q) = I_A(\inputdfs,q,\outputdf) - I_A(\inputdfs-R,q,\outputdf')\] 
\end{definition}

Namely, we remove the rows set $R$ from the input dataframes $\inputdfs$, employ again the operation $q$ which now results in a new output dataframe $\outputdf'$, and recalculate the interestingness score $I_A$.  
Intuitively, the higher the decrease in the interestingness score caused by removing $R$, the higher its contribution is to the interestingness of the column $\outputdf[A]$.

\begin{example}
\label{example:contribution}
Recall that Example~\ref{example:interestingness} showed that the most interesting column in the filter step (Figure~\ref{fig:spotify_filter}) is `decade', obtaining a score of $0.56$.
Calculating, for example, the contribution of the `decade'=``2010s'', we first omit from $\inputdf$ all rows where `decade'=``2010s''.
We then employ the operation $q$, i.e., \textit{``filter by `popularity' greater than 65''}) and recalculate the interestingness. 
The resulted score is now $0.47$, which means that the contribution score of ``2010s'' is $C(R_{\text{``2010s''}},A_{\text{decade}},Q)=0.086$.
This is a relatively high contribution score, as the interestingness of the column `decade' decreased by 16\% when removing songs made in the 2010s. 


\end{example}

Note that generally, the contribution of a set-of-rows may be negative, if removing this set-of-rows reduces the value of the interestingness function. 
For example, for group-by queries with diversity as an interestingness function, the contribution function may either be negative or positive. To see this, consider the  dataframe $\inputdf = \{(x,1), (x,2), (y,3)\}$ and the group-by query that groups the first attribute and sums the second. Here $\outputdf = \{(x,3), (y,3)\}$ (diversity $=0$), whereas if we remove $(x,2)$ from $\inputdf$ and perform the same query, we will get $\outputdf = \{(x,1), (y,3)\}$ (diversity $>0$). Thus, the contribution of $(x,2)$ is negative. 
However, if we consider the dataframe $\inputdf = \{(x,1), (x,1), (y,1)\}$ with the same query, we will get $\outputdf = \{(x,2), (y,1)\}$ (diversity $>0$). If we remove one of the $(x,1)$ tuples and evaluate the same query, we get $\outputdf = \{(x,1), (y,1)\}$ (diversity $=0$). Thus, the contribution of $(x,1)$ is positive. 
As we describe below, we are interested in sets-of-rows that obtain a significantly high, positive contribution score, compared to other sets-of-rows. 
If there are no sets-of-rows with a positive contribution, no explanation will be generated for the exploration step (see Section \ref{sec:skyline}).

\subsection{Explanation Candidates}
As mentioned above, \sys{} \textit{explains} a given exploratory step by identifying sets-of-rows in its \textit{input} dataframes that contribute to the interestingness score $I_A(Q)$, on column $A$ in the \textit{output dataframe}.
An explanation-candidate is defined as follows:

\begin{definition}[Explanation Candidate]
For exploratory step $Q=(\inputdfs,q,\outputdf)$,
an explanation-candidate $E$ is defined by $E \coloneqq (R,A)$ where $R \subset \inputdfs$ and $A \in \mathcal{A}(\outputdf)$
\end{definition}


\begin{example}
Consider again the filter step depicted in Figure~\ref{fig:spotify_filter} resulting in songs with popularity over 65. An example explanation-candidate is $(R_{'decade'=2010s'},A_{'decade'})$,
namely the rows from $\inputdf$ (i.e., the full songs dataframe) depicting songs released in the 2010s decade, and the `decade' column in the output dataframe.
To understand how $(R_{'decade'=2010s'},A_{'decade'})$ explains the filter step, recall that the interestingness score here measures the deviation from $\inputdf$ to $\outputdf$. As calculated in Example~\ref{example:contribution}, the contribution $C(R_{'decade'=2010s'},A_{'decade'},Q)$ is high, which means that songs from 2010s 
\textit{explain} this deviation. Indeed, as also illustrated in the final explanation (figure~\ref{fig:explained_filter}),  \textit{songs (rows) from 2010s are highly more frequent \textit{after} the filter, than in the entire dataset}.   
\end{example}

Naturally, not all explanation-candidates are useful. We next explain how we restrict the set-of-rows to include only the ones that are semantically related, then describe the quality metric of the explanation-candidates and how \sys{} only selects the best ones and return them as coherent captioned visualizations.

\subsection{Partitioning the Input Dataframe}\label{sec:binning} 
While, in theory, one could calculate the contribution of individual rows, or, alternatively, do so for \textit{all} sets-of-rows, \sys{} focuses on the contribution of \textit{semantically related sets-of-rows}. This both allows for a faster computation (as is shown in the sequel) and yields explanations that depict the ``bigger picture'', allowing users to obtain meaningful, high-level insights from each exploratory step. 
The following example illustrates the  latter point. 




\begin{example} 
Consider again the group-by results in Figure~\ref{fig:spotify_groupby}. Recall from Example~\ref{example:interestingness} that the column `loudness' obtains high interestingness score, as it contains a diverse set of values. 
Had we computed the contribution of, e.g., each single row in $\inputdf$, we would have seen that the top-2 contribution values are of `1991' and `2007', since they have the most extreme values (-11.07 and -7.49, resp). 
Yet instead, if we group together years by their corresponding decade, we now obtain that the highest contribution is made by the decade `1990s'. This now yields a higher-level interesting pattern indicating that \textit{songs released in the 1990s tend to be \underline{less loud} than in later decades}. As detailed in the sequel, this observation will be highlighted by the output explanation of \sys{} (See Figure~\ref{fig:explained_groupby}).
\end{example}

We next define the row-partition scheme and detail the specific methods currently implemented in \sys{}. 


\begin{definition}[Row Partition]
\label{def:partition}
Given an input dataframe $\inputdf \in \inputdfs$, 
a row partition divides $\inputdf$ into $n+1$ disjointed sets-of-rows: $\mathcal{R} = \{R_1,R_2,\dots,R_n,\hat{R}\}$
such that: 
\begin{equation*}
\forall R_i,R_j \in \mathcal{R},~~~ R_i \subset \inputdf \wedge R_j \subset \inputdf \wedge \bigcup_{R_i \in \mathcal{R}} R_i \cup \hat{R} = \inputdf \wedge R_i \cap R_j = \emptyset     
\end{equation*} 
The special set-of-rows $\hat{R} \in \mathcal{R}$ (can be empty) is called an \textit{ignore-set}, as it cannot become an explanation-candidate, as detailed below. 
\end{definition}

\sys{} supports three types of row partition methods and utilizes them when generating explanation (as described in Section~\ref{sec:skyline}). Other partition methods are supported, as long as they comply with definition~\ref{def:partition}. The partition methods in \sys{} are as follows:






\vspace{1mm}
\paratitle{Frequency-based partition}
Given either a numeric or categorical attribute $A$ we divide $\inputdf$ to $n$ sets-of-rows, corresponding to the $n$ most prevalent values in the column $\inputdf[A]$. We then assign the rest of rows in the ignore set $\hat{R}$.

\paratitle{Numeric-based partition} Given 
a numeric attribute $A$, we divide the rows in $\inputdf$ according to their corresponding values in $\inputdf[A]$ to $n$ sets-of-rows, using \textit{equal-frequency binning}.
Namely, each set-of-rows in this partition correspond to an interval of values of $\inputdf[A]$, s.t. the number of values in each interval is equal. The ignore-set $\hat{R}$ is empty in this case. 


\paratitle{Many-to-one partition} 
This method partitions the rows by mining many-to-one relationships between columns in $\inputdf$. 
Given an attribute $A$, we look for a different attribute $B$
s.t. each value in $\inputdf[A]$ is mapped to a single value in $\inputdf[B]$, yet there exist at least two values in $\inputdf[B]$ mapped to different values of $\inputdf[A]$. Formally, for an attribute $A$ in $\inputdf$ we search for all columns $B$ such that both of the following conditions hold:
\begin{enumerate}
\item\label{itm:first} $\forall r_i, r_j \in \inputdf, (r_i[A]=r_j[A]) \rightarrow (r_i[B]=r_j[B])$
\item\label{itm:second} $\exists r_i,r_j \in \inputdf, (r_i[B]=r_j[B]) \wedge (r_i[A]\neq r_j[A])$ 
\end{enumerate}
Intuitively, these conditions ensure that the partition of values from $A$ according to column $B$ are consistent, i.e., every pair of equal values is placed in the same set (Condition \ref{itm:first}), and that the partition according to $B$ will be strictly coarser compared to the partition according to $A$ (Condition \ref{itm:second}). 

After mapping the values from $A$ to $B$ we split the set-of-rows using the frequency-based partition (as defined above), over the column $B$.
This partition is particularly useful for group-by dataframes, as exemplified below.

\begin{example}
Consider again the group-by output dataframe in Figure~\ref{fig:spotify_groupby}, where each row represents a single year. Partitioning the rows via the `year' column could be done using the frequency-based method, or many-to-one (as the column is categorical, numeric binning is not applicable). 
Using the frequency-based method, we can partition the rows in the input dataframe (i.e., before applying the group-by operation) according to the $n$ most frequent `year' values and placing the rest of the rows in the ignore-set $\hat{R}$. In the many-to-one partition, we identify that the column `decade' has a many-to-one relationship with `year'. This method has yielded a preferable explanation, particularly for the set-of-rows associated with  \textit{`decade'=`1990s'}.
\end{example}




\subsection{Quality of Explanation Candidates}
\label{sec:candidate}

Intuitively, an explanation-candidate $E=(R,A)$ is a good explanation for the exploratory step $Q$
if the contribution $C(R,A,Q)$ to the interestingness of $A$ is significant, and the interestingness score $I_A(Q)$ is itself high.  
We next explain how the significance of contribution is calculated, and how \sys{} balances between interestingness and contribution. 

We evaluate the significance of contribution for a given set-of-rows $R \in \mathcal{R}$ as follows. Rather than considering the raw contribution score of a set-of-rows $R$, we compare the contribution of $R$ to the contributions of the other sets-of-rows in $\mathcal{R}$.
We then define the \textit{standardized contribution} of a set-of-rows $R \in \mathcal{R}$ by: 
\begin{equation*}
    \Bar{C}(R,A) = \frac{C(R,A,Q) - \mu_{\mathcal{R}}}{s_{\mathcal{R}}}
\end{equation*}
Where $\mu_\mathcal{R}$ and $s_\mathcal{R}$ are the mean and standard deviation of the contribution scores of all sets-of-rows in the partition $\mathcal{R}$.

This quantifies the significance of the contribution of $R$, since the higher the \textit{standardized} contribution, the farther the contribution of $R$ from the mean contribution of its fellow row sets.

Correspondingly, the \textit{quality} of an explanation $(R,A)$ is measured by using two metrics: (1) the standardized contribution $\Bar{C}(R,A)$ and (2) the interestingness of the column $A$, $I_A(Q)$.



\paratitle{Skyline of contribution \& interestingness} We next define the set of desired explanations based on contribution and interestingness. Denote by $EC(Q)$ the set of all explanation-candidates considered in \sys{} for a step $Q$ is, formally:
$$EC(Q) \coloneqq \bigcup_{\mathcal{R}}\bigcup_{R \in \mathcal{R}}\bigcup_{~A \in \mathcal{A}(\outputdf)}\left\{( R,A) \right\}$$


Given the set $EC(Q)$ of all explanation-candidates, we look for ones that obtain \textit{both} a good contribution and a high interestingness score. To balance the two metrics we use a skyline-operator~\cite{borzsony2001skyline} calculation. Namely, we define the set $EX$ of desired explanations as a maximal subset of $EC(Q)$ satisfying the following: 
\begin{equation*}
\begin{split}
    \forall (R,A) \in EX.~~\nexists (R',A') \in EC(Q).~  (I_{A'}(Q) > I_A(Q) \land \\ \bar{C}(R',A') > \bar{C}(R,A))
\end{split}
\end{equation*}


\begin{example}
\label{example:skyline}
Reconsider the group-by step, depicted in Figure~\ref{fig:spotify_groupby}. Recall from Example~\ref{example:interestingness} the interestingness of column `loudness' is higher than of `danceability'.
The set-of-rows with the highest contribution is the one where `decade'=``1990s'' (when using the many-to-one partition), obtaining raw influence score of $1.12$, whereas the sets of rows associated with ``2000s'', ``2010s'', ``2020s'' obtained contribution of $-0.04$,$-0.35$, and $-0.055$ (resp).
Therefore, ``1990s'' obtains the highest \textit{standardized} contribution of $1.69$. This explanation-candidate is indeed a \textit{dominating} one, according to the skyline computation, and therefore returned to the user by \sys{}. 
\revb{Note that the skyline outputs another dominating explanation, having a \textit{lower} interestingness score, yet a \textit{higher} standardized contribution. This is `decade'=``2020'' w.r.t. interestingness $I_{danceability}$, which has an interestingness score of $0.04$ (as is shown in Example~\ref{example:interestingness}), and a standardized contribution of $1.7$. According to this explanation (visualization omitted), \ul{\textit{Songs made in the 2020's are relatively more ``danceable'' than older songs} }.}

\end{example}

\subsection{Explanations Generation Process}
\label{sec:skyline}

\SetCommentSty{rmfamily}
 \IncMargin{1em}
 \begin{small}
 \begin{algorithm}[t]
\DontPrintSemicolon
   \KwIn{Exploratory step $Q=(\inputdfs,q,\outputdf)$}
   \KwOut{Explanations for $Q$}
    
    
    \label{ln:partial}  \For{$A \in \mathcal{A}(\outputdf)$}{
        Calculate interestingness score $I_A(Q)$
    }
    
     $\mathcal{SR} \leftarrow \emptyset$\;
     
     \label{ln:for}  \ForEach{ row-partition}{
        $\mathcal{R} \leftarrow \textit{row-partition}(\inputdfs)$\;
       
         $\mathcal{SR} \leftarrow \mathcal{SR} \cup \{\mathcal{R}\}  $\;
    }
     $EC  \leftarrow$ Empty Dictionary\;
    
    \label{ln:for}  \ForEach{ $\mathcal{R} \in \mathcal{SR}, A \in \mathcal{A}(\outputdf)$}{
         \ForEach{ $R \in \mathcal {R}$}
        {
         Calculate contribution $C(R,A,Q)$\;
            \If{$C(R,A,Q) > 0$}
            {
            $EC[(R,A)] \leftarrow \left(\Bar{C}(R,A), I_A(Q) \right)$\;
            }
        }

    }
    $EX = \underset{(R,A) \in EC}{\operatorname{SKYLINE}} \left(\Bar{C}(R,A), I_A(Q)\right)$\;
    
    \ForEach{$E \in EX$}
    {
        GenerateVisualExplanation($E$)\;
    }

\caption{\sys{} Explanations Generation}
 \label{alg:alg}
\end{algorithm}
\end{small}

The explanations generation process for an exploratory step $Q$ is detailed in Algorithm~\ref{alg:alg}. 
It takes as input an exploratory step $Q=(\inputdfs,q,\outputdf)$ and operates as follows.

\paratitle{Pre-processing: interestingness \& row partitioning}
First, depending on the type of operation $q$ (filter, group-by, join) we calculate the interestingness scores $I_A(Q)$ using the corresponding function, for each attribute $A$ in the output dataframe $\outputdf$ (Lines 1--2 in the algorithm).
Next, we use our row partition techniques (Section~\ref{sec:binning}) to split the input dataframes into multiple partitions $\mathcal{R}_1,\mathcal{R}_2,\dots$, and unify all sets-of-rows into the set $\mathcal{SR}$ (Lines 3--6).

\paratitle{Forming explanation-candidates and calculating standardized contribution}
We initialize an empty dictionary $EC$ (Line 7), which will contain the quality scores for each explanation candidate. 
Then, we iterate over all partitions and all output attributes (Line 8). For each pair of partition and attribute, we iterate over every set-of-rows in the partition (Line 9) and compute its contribution to $A$ (Line 10). If the contribution is positive (Line 11), then both the standardized contribution (w.r.t. the partition, as explained in Section~\ref{sec:candidate}), and the interestingness score of $A$ are stored in $EC[(R,A)]$ (Line 12).

\paratitle{Calculating the interestingness/contribution skyline}
Now that the dictionary $EC$ contains for each explanation candidate $(R,A)$ its (standardized) contribution and interestingness score, we employ the skyline operator (see Section~\ref{sec:candidate}) which outputs only the dominating explanations, that are not inferior to any other candidates in both contribution or interestingness (Line 13). 

\revb{Note that to further limit the resulted explanations,
one can use, e.g., a weighted average between the interestingness score and the standardized contribution. 
Namely, given user-defined weights $W_I$ and $W_C$, let the weighted score of an explanation candidate be  $SCORE(R,A) = \frac{W_I\cdot I_A(Q) + W_C\cdot \bar{C}(R,A) }{W_I + W_C}$. We can then use this score to rank the explanations generated by the skyline operator (Line 13) and keep the top $k$ ones. }





\paratitle{Generating captioned visualizations for each resulted explanation}
For each dominating explanation $(R,A)$, s.t. $R \in \mathcal{R}$ we now (Lines 14--15) generate a corresponding captioned visualization (as illustrated in Figure~\ref{fig:explained_new}).
\sys{} produces a different visualization for each type of interestingness measure:
(1) for exceptionality-based explanations, we highlight the deviation in column $A$, caused by the filter/join operation. This is done, as illustrated in Figure~\ref{fig:spotify_filter}, via a side-by-side bar plot, in which the left-hand side depicts the mean $\inputdf[A]$ value, for all sets-of-rows in the partition $\mathcal{R}$, and the right-hand side depicts the mean $\outputdf[A]$. The chosen set-of-rows for the explanation, $R$, is colored in green. As for the caption, we use a natural language template and plug-in the attribute name $A$, and the label of $R$ according to the partition method. The label is set based on the partition approach (Section \ref{sec:binning}). If the chosen partition is numeric-based, the end values of the interval are set as the label. If the partition is many-to-one, the value in column $B$ will be the label. Otherwise, the partition will be frequency-based, and the label is the value itself. 
We then describe the deviation of $R$ from $\inputdf[A]$ to $\outputdf[A]$ in percentages and multiplications, as illustrated in Figure~\ref{fig:explained_filter}.
(2) Diversity-based explanations highlight the extremity in terms of the aggregated $A$ values in $\outputdf[A]$ obtained by the rows in $R$, compared to the rest of the sets-of-rows in the partition $\mathcal{R}$. As illustrated in Figure~\ref{fig:spotify_groupby}, we use a bar chart to depict the mean aggregated values obtain by each set-of-rows in $\mathcal{R}$, where the mean value of the rows in $R$ is again colored in green. To further emphasize the extremity of $R$, we also depict the mean $\outputdf[A]$ value, using a horizontal red line. The caption is generated in a similar manner to that of the exceptionality-based explanations, but supports the visualization by emphasizing how far is the value of $R_j$ from the mean, in terms of standard deviations.

\paratitle{Sampling optimization}
To reduce explanations generation times, we employ a sampling approach to optimize Algorithm 1.  
Instead of considering all rows in $\inputdf$ (Lines 1--2), we calculate the interestingness scores over a sample of the rows, obtained using uniform sampling. All other parts of the algorithm stay intact, and, in particular, the contribution is still computed over all rows.
In Section \ref{sec:experiments}, we show that with a relatively small sample size (5K), this approach achieves good accuracy while substantially improving interactivity.  

\reva{
\subsection{Customization \& Extensions of \sys}
\label{sec:generalization}
We next detail several extensions for \sys.

\paratitle{General interestingness functions}
While \sys{} utilizes established interestingness measures shown useful in exploration tasks~\cite{van2010maximal,VartakRMPP15,bar2020automatically,milo2019predicting}, \sys\ can take any interestingness function as input, with no required properties (e.g., monotonicity or non-negativity).
Examples measures are compactness/coverage~\cite{chandola2007summarization} for group-by operations, and \textit{surprisingness}~\cite{liu1999finding}, as well as learned-based measures (inspired by~\cite{luo2018deepeye}).

 


\paratitle{\shep{User-specified} columns}
Furthermore, to give expert users more control, they can specify the columns that they are interested in.
Then, \sys\ will only compute the skyline explanations for the chosen columns based on their interestingness and the contribution of the different sets-of-rows to these columns. 
This is simply done by projecting the input and output dataframes over the user-selected set of attributes, before employing Algorithm~\ref{alg:alg}.

For instance, to explain the filter step in our running example (See Figure~\ref{fig:spotify_filter}) the user could potentially restrict \sys\ to, e.g., the `danceability' and `loudness' columns (rather than on all columns), hence obtaining explanations regarding these columns only.  

\paratitle{Custom partitioning of rows}
\reva{
Last, as mentioned above, our framework can be further extended with a user-defined partitioning scheme, as long as they comply with Definition~\ref{def:partition}. 
For instance, users can add a custom partition for date/time columns and group the tuples by month or years; partition a geo-location column by city or state, etc. The new partitions are added to the existing ones, and the system uses them all to generate explanations.  }
}

%% file: experiments.tex
\section{experiments}\label{sec:experiments}
We evaluated the quality and performance of \sys\ on three real-world datasets, and compared them to existing baselines. The results show that the hybrid explanations generated by \sys\ and its sampling optimization version, \sysopt, are clearer and more interesting than the ones generated by our baselines, and that the sampling optimization employed in \sys\ allows to generate explanations in interactive time while not significantly reducing the accuracy of obtained explanations w.r.t. the skyline. 


\subsection{Setup, Datasets, Queries, and Baselines}
Our experimental study includes 3 real-world datasets and 4 baselines that are based on expert knowledge and previous work. 

\paratitle{Implementation}
\sys\ is implemented in Python 3.8. It uses Pandas \cite{reback2020pandas} to store and manipulate the database and uses NumPy \cite{harris2020array} to compute the explanations and Matplotlib \cite{matplotlib} to generate the visualizations. We have made the source code available \cite{repo}. 
The experiments were run on Windows 10 laptop with 16GB RAM and 1.9 GHz Quad-Core Intel Core i7 processor.

\vspace{1mm}
\paratitle{Datasets}
We have used the following 3 datasets:

\noindent{\bf 1. Spotify} \cite{spotifydataset}: containing information about tracks (e.g., duration, popularity, danceability) and artists (e.g., genres and popularity). It comprises a single table of $174,389$ rows and $20$ columns.

\noindent {\bf 2. Credit Card Customers} \cite{creditdataset}: containing customer details such as age, gender, education level, and credit card category. It has a single table with $10,127$ rows and $21$ columns.

\noindent {\bf 3. Products and Sales} \cite{products}: consisting of two tables: a Products table ($9,977$ rows, $16$ columns) with information about beverage products (e.g., name, vendor, price), and a Sales table ($3,049,913$ rows and $17$ columns) that contains a record of all the sales of products in a given chain (e.g., store id, quantity, date). The join view of the two tables has $3,049,913$ rows. For our scalability experiments  (Section~\ref{sec:automatic-exp})
we uniformly sampled additional $6,950,087$ rows (i.e., duplicates) to get a view with exactly 10M rows.

\shep{
Further note that the datasets all contain skewed columns. In total, over 31 of the columns are highly skewed, and over 41 of the columns are moderately skewed. 
In particular, Fisher-Pearson standardized moment coefficient \cite{brown2011measures} was high for multiple columns in each dataset. For example, the top-1 column had a measurement of 10.16, 2.06, and 205.89 for the Spotify, Credit Card Customers, and Products and Sales datasets, respectively.
}



\paratitle{Queries} We have composed 5 filter/join queries for each dataset, and 5 group-by queries for each dataset. 
The queries, along with their reference numbers, are shown in Tables \ref{tbl:queries} and \ref{tbl:gb_queries} (\cref{sec:queries}), where Credit Card Customers is referred to as `Bank'. 

\begin{figure*}[t]
    \centering
    \begin{subfigure}{1\linewidth}
    \centering
    \includegraphics[scale=0.15]{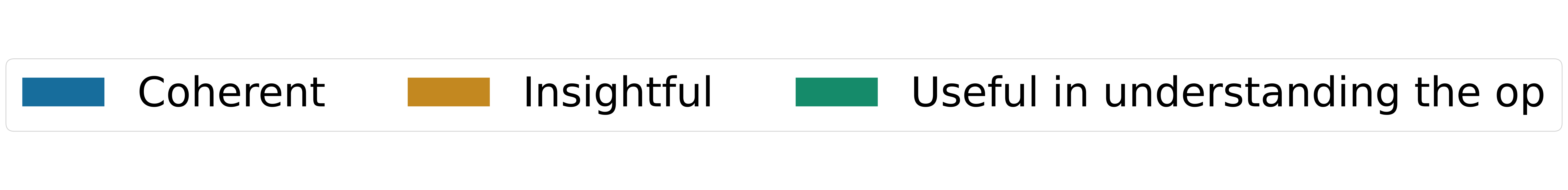}
    \vspace{-8mm}
    \caption*{} 
    \label{fig:user_study_legend}
    \end{subfigure}
    \begin{subfigure}{.33\linewidth}
    \includegraphics[scale=0.3]{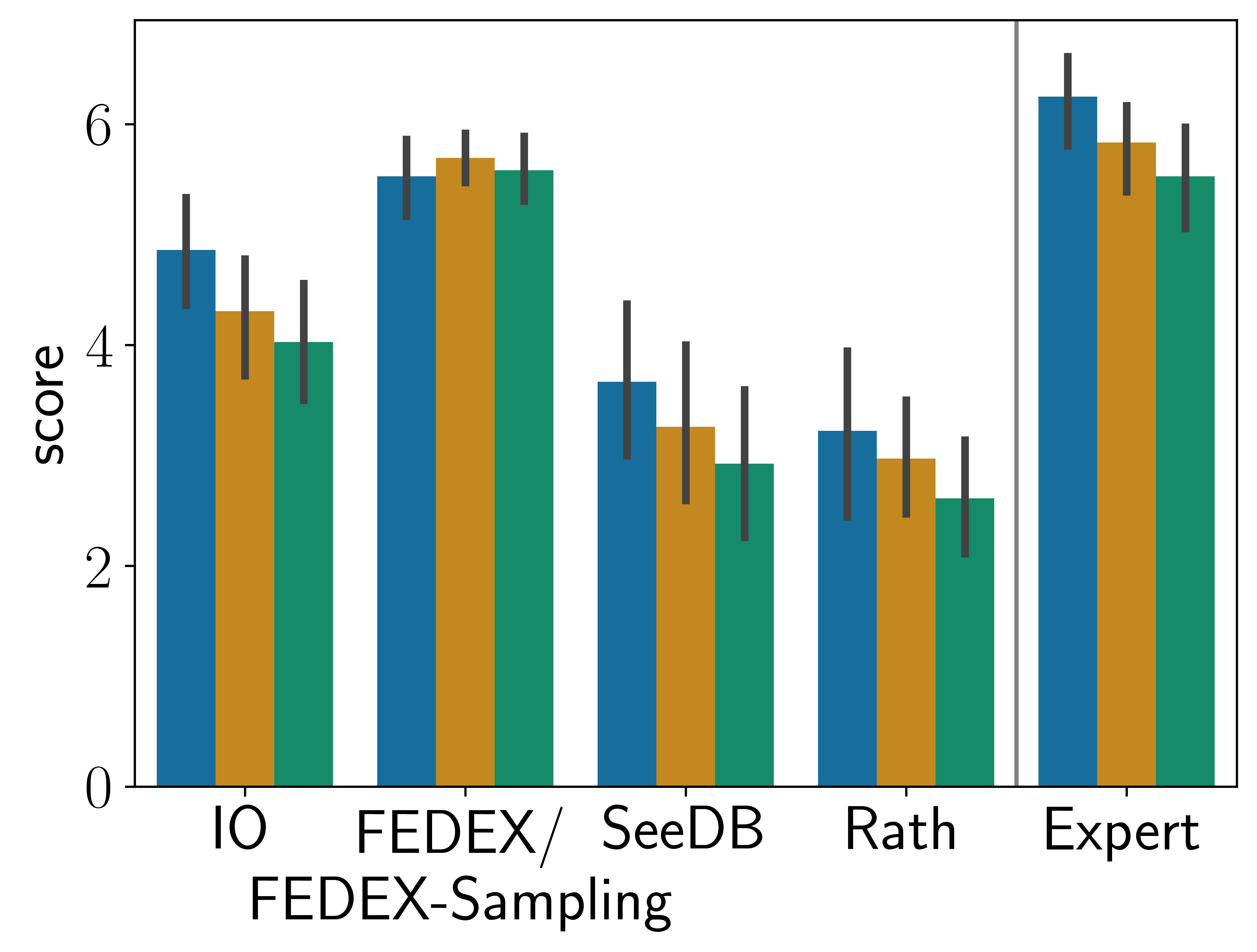}
    \caption{\revc{Results for the Credit Card dataset}}
    \label{fig:user_study_bank}
    \end{subfigure}%
    \begin{subfigure}{.33\linewidth}
    \centering
    \includegraphics[scale=0.3]{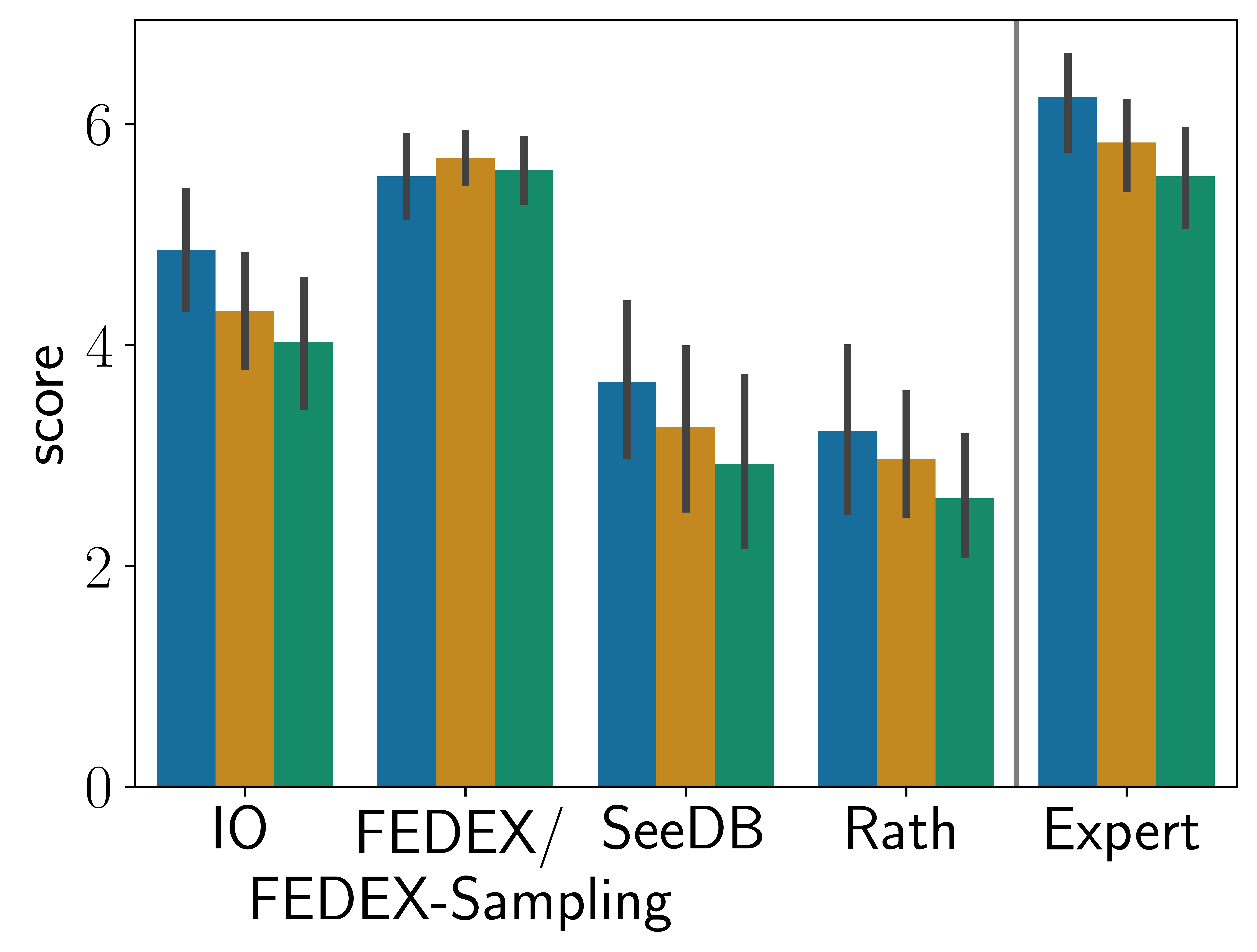}
    \caption{\revc{Results for the Spotify dataset}}
    \label{fig:user_study_spotify}
    \end{subfigure}%
    \begin{subfigure}{.33\linewidth}
    \centering
    \includegraphics[scale=0.3]{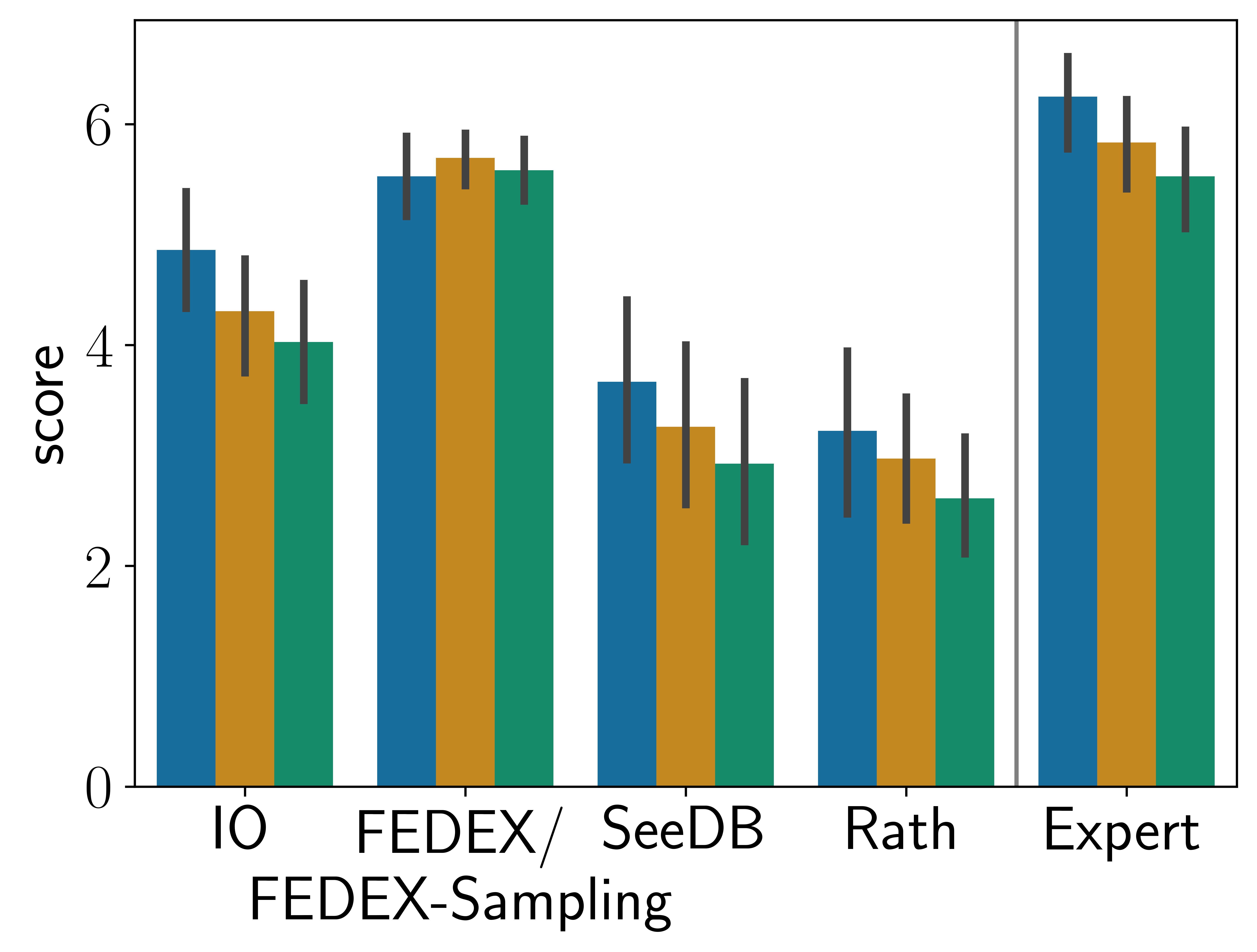}
    \caption{\revc{Results for the Products dataset}}
    \label{fig:user_study_products}
    \end{subfigure}
    \caption{\revc{User study results on the three datasets}}
    \label{fig:user_study}
\end{figure*}

\begin{figure}[h]
    \centering
    \includegraphics[scale=0.4]{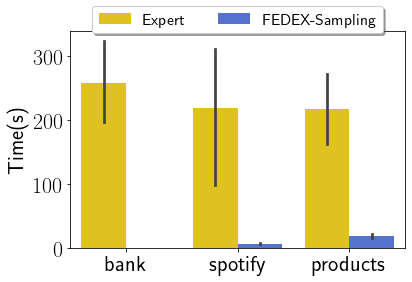}
    \caption{Explanation generation time for the user study in Fig. \ref{fig:user_study}}
    \label{fig:expert_gen_time}
\end{figure}

\begin{figure}[h]
    \centering 
    \includegraphics[width=2.2in]{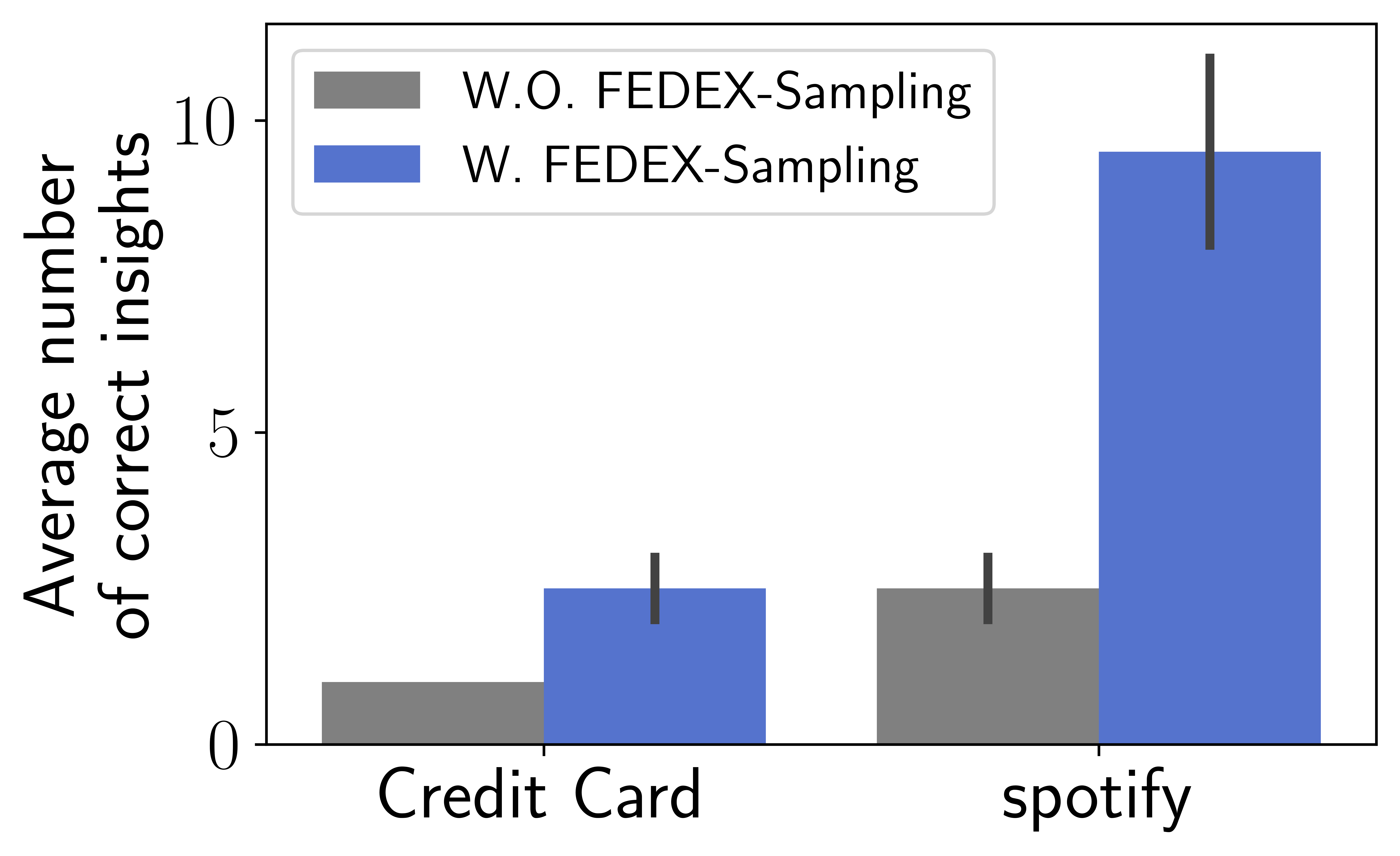}
    \caption{\reva{Interactive user study}}
    \label{fig:user_study_insights}
\end{figure}

\begin{figure}[h]
    \centering 
    \includegraphics[scale=0.35]{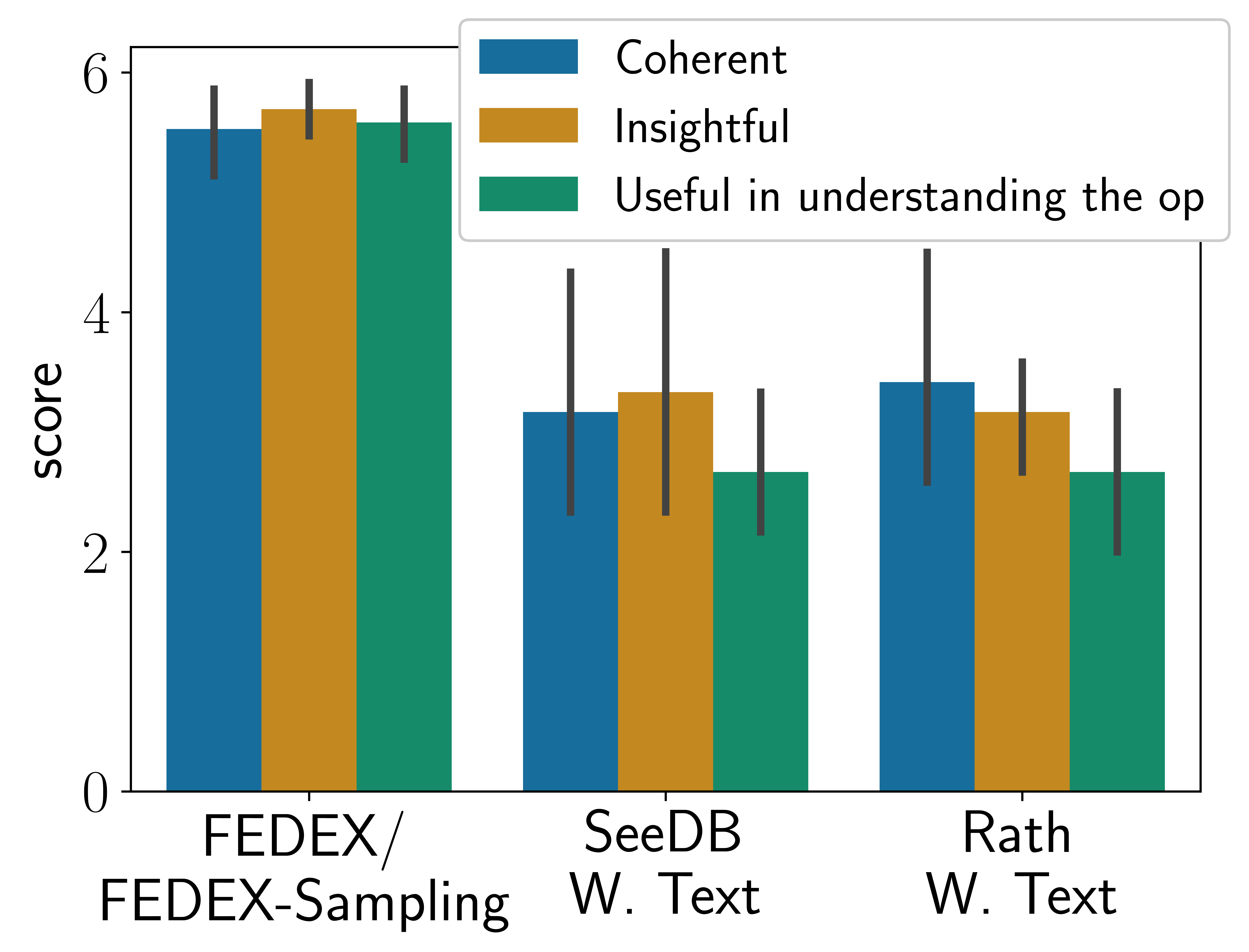}
    \caption{\shep{User study results for baselines augmented with NL explanations}}
    \label{fig:NL_baseline_userstudy}
\end{figure}

\paratitle{Baselines and optimized version} 
Our experiments included a comparison of \sys\ to the following alternatives:

    \noindent {\bf 1. \seedb} \cite{VartakRMPP15}: The solution of Vartak et. al. (implemented in \cite{seedbrepository}) automatically generates visualizations that are meant to emphasize the interesting trends in a given view stemming from a query over a source view.
    
    \noindent {\bf 2. \rath} \cite{tang2017extracting}: RATH (implemented in implementation is taken from \cite{rath}) automatically generates the top-k insightful visualizations using a single score function for all operations that is both applicable to different types of insights and fair across different types of insights (for a detailed comparison see Section \ref{sec:related}). 
    
    \noindent {\bf 3. \io}: The {\bf I}nterestingness {\bf O}nly baseline is based on previous work \cite{0002M13} where the influence of an attribute is measured by checking the difference in interestingness of an attribute in $\outputdf$ w.r.t. $\inputdfs$. 

    \noindent {\bf 4. \expert}: As part of our user study (Section \ref{sec:user-study}), we have asked three experts to manually formulate their own explanations  for each of the three datasets, each represented in its own notebook. These explanations consisted of a detailed textual description. The experts had access to the 
    query, the dataset, and the resulting dataframe. 

\noindent {\bf 5. \sysopt}: The optimized version of \sys, where uniform sampling of the rows is used to compute the interestingness. 
    After performing thorough experiments to choose a reasonable sample size that does not compromise the accuracy of the generated explanations (See the paragraph titled ``Accuracy of \sysopt'' in Section \ref{sec:automatic-exp}), we have concluded that a sample of 5K rows delivers good accuracy.

\begin{figure*}[t!]
    \centering
    \begin{subfigure}[b]{.33\linewidth}
    \includegraphics[scale=0.5]{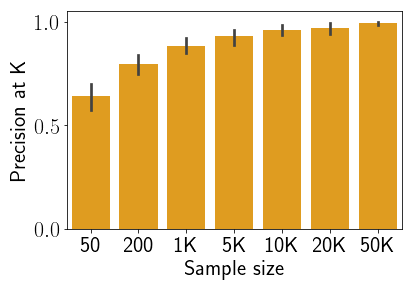}
    \caption{Precision@k for \sysopt}
    \label{fig:precision_at_k}
    \end{subfigure}%
    \begin{subfigure}[b]{.33\linewidth}
    \centering
    \includegraphics[scale=0.5]{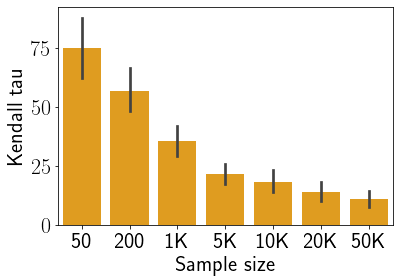}
    \caption{Kendall-Tau distance for \sysopt} 
    \label{fig:kendall_tau}
    \end{subfigure}
    \begin{subfigure}[b]{.33\linewidth}
    \centering
    \includegraphics[scale=0.5]{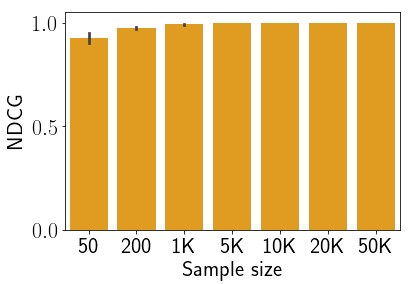}
    \caption{nDCG for \sysopt}
    \label{fig:ndcg}
    \end{subfigure}
    \caption{Accuracy results for \sysopt\ averaged over the Spotify and Products datasets with all relevant filter and join queries in \cref{tbl:queries}}
    \label{fig:accuracy_samples}
\end{figure*}

\begin{figure*}[t!]
    \centering
    \begin{subfigure}[b]{.33\linewidth}
    \includegraphics[scale=0.38]{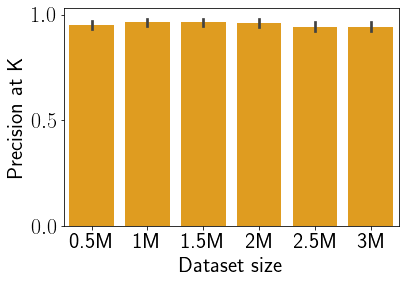} 
    \caption{\revc{Precision@k for \sysopt}}
    \label{fig:precision_at_k_rows}
    \end{subfigure}%
    \begin{subfigure}[b]{.33\linewidth}
    \centering
    \includegraphics[scale=0.38]{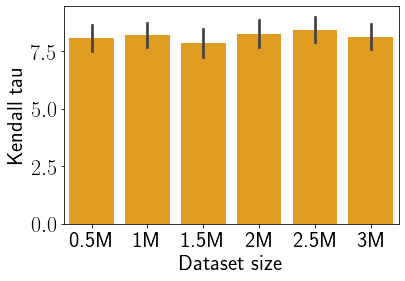}
    \caption{\revc{Kendall-Tau distance for \sysopt}} 
    \label{fig:kendall_tau_rows}
    \end{subfigure}
    \begin{subfigure}[b]{.33\linewidth}
    \centering
    \includegraphics[scale=0.38]{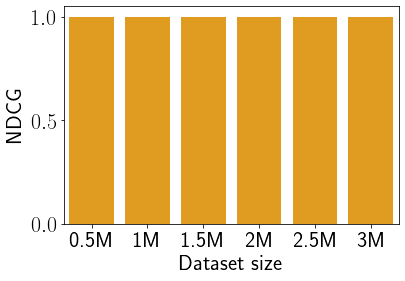}
    \caption{\revc{nDCG for \sysopt}}
    \label{fig:ndcg_rows}
    \end{subfigure}
    \caption{{\revc{Accuracy results for \sysopt\ for the Products datasets with all relevant filter and join queries in \cref{tbl:queries}}}} 
    \label{fig:accuracy_rows}
\end{figure*}

\begin{figure*}[t!]
    \centering
    \begin{subfigure}[b]{.33\linewidth}
    \includegraphics[scale=0.33]{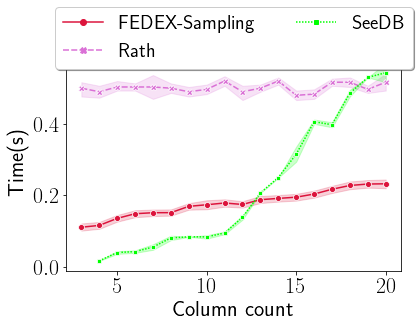}
    \caption{\revc{Credit Card Customers dataset}}
    \label{fig:runtime_columns_bank}
    \end{subfigure}%
    \begin{subfigure}[b]{.33\linewidth}
    \centering
    \includegraphics[scale=0.33]{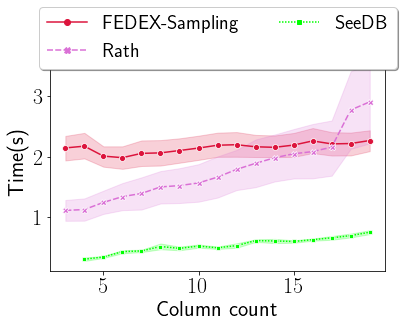}
    \caption{\revc{Spotify dataset}} 
    \label{fig:runtime_columns_spotify}
    \end{subfigure}
    \begin{subfigure}[b]{.33\linewidth}
    \centering
    \includegraphics[scale=0.33]{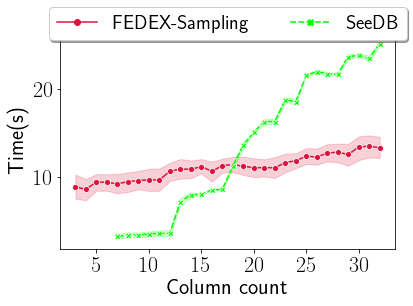}
    \caption{\revc{Products and Sales dataset}}
    \label{fig:runtime_columns_products}
    \end{subfigure}
    \caption{\revc{Runtime as a function of column number for all three datasets for \sysopt\ and the baselines averaged over the filter and join queries from \cref{tbl:queries}
    }}
    \label{fig:runtime_columns}
\end{figure*}

\begin{figure*}[t!]
    \centering
    \begin{subfigure}[b]{.33\linewidth}
    \includegraphics[scale=0.3]{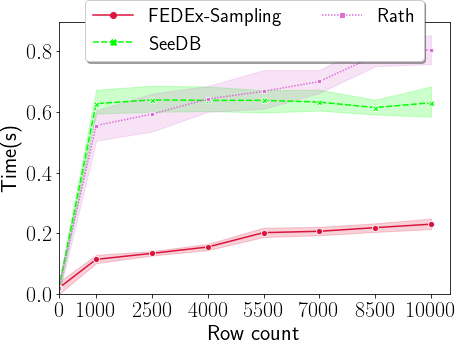}
    \caption{\revc{Credit Card Customers dataset}}
    \label{fig:runtime_rows_bank}
    \end{subfigure}%
    \begin{subfigure}[b]{.33\linewidth}
    \centering
    \includegraphics[scale=0.3]{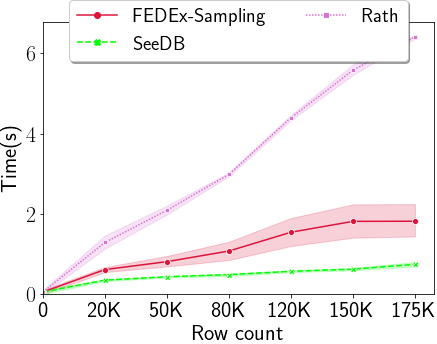}
    \caption{\common{Spotify dataset}} 
    \label{fig:runtime_rows_spotify}
    \end{subfigure}
    \begin{subfigure}[b]{.33\linewidth}
    \centering
    \includegraphics[scale=0.3]{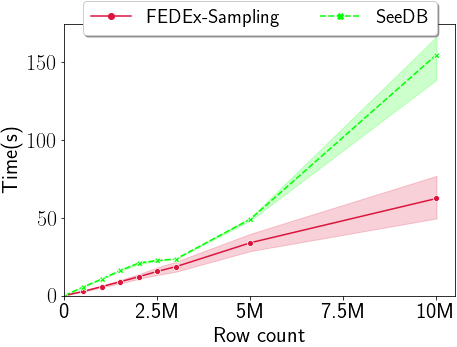}
    \caption{\common{Products and Sales dataset}}
    \label{fig:runtime_rows_products}
    \end{subfigure}
    \caption{\shep{Runtime as a function of row number for all three datasets for \sys\ and \sysopt\ averaged over the filter and join queries from \cref{tbl:queries}}}
    \label{fig:runtime_rows}
\end{figure*}

\subsection{User Studies}\label{sec:user-study}
\common{We detail two user studies that were performed to evaluate the quality of the explanations generated by \sys\ and \sysopt. 
For our user studies, we have used separate notebooks for each dataset and the appropriate queries.
The maximal number of explanations presented to users (the size of the skyline set) over all queries in both studies was 2. 
} 

\paratitle{Comparison to existing baselines}
We have performed a user study with the goal of evaluating the quality of the explanations generated by \sys\ compared to existing baselines and expert-generated explanations. that included 25 participants. The users had different backgrounds; 17 of the users were non-experts with minimal SQL knowledge and 8 were CS graduate students. 
The study was based on three notebooks, one for each of the three datasets. The notebooks also included details regarding the given dataset and the goal of the exploration session (the notebooks can be found in the \sys\ repository \cite{repo}). 
The Spotify notebook contained the filter/join queries 6, 7 and the group-by queries 21, 22 (see \cref{sec:queries}); 
the Credit Card Customers notebook contained the filter/join queries 11, 12, 13 and group-by query 27 from \cref{tbl:gb_queries}; 
and the Product and Sales notebook contained the filter/join queries 1, 5 and group-by queries 16, 17, 18 from \cref{tbl:gb_queries}. 
The Spotify and Credit Card notebooks were each shown to 9 users, and the notebook with the Products and Sales session was shown to 10 users. 
Each notebook also included the explanations generated by \seedb, \rath, \io, \expert, and \sys\ \revc{(the explanations computed by \sysopt\ were identical to those computed by \sys, i.e. the skyline set was identical)}. 
For the \expert\ baseline, we have asked 3 experts to analyze the notebooks and generate an explanation for each operation manually. Figure \ref{fig:expert_gen_time} shows the time it took the experts to generate the explanations compared to the generation time of \sys.
Naturally, the explanation generation time of the experts was substantially larger than for \sys.

For each user, we presented the query, the original input dataset and the query output. Then, we showed the user up to five explanations (the \seedb\ baseline could not generate explanations for group-by queries as it compares $\inputdf$ and $\outputdf$, but in group-by operations the input and output columns are different). 
Users were asked to grade the explanations on a scale of 1--7 w.r.t. \textit{coherency} (is the explanation easy to understand?), \textit{insight level} (does the explanation provide an interesting insight?), and \textit{usefulness} in understanding the operation (does the explanation assist in understanding the EDA operation results?).

For all three datasests, we observed that users have generally preferred the \expert\ explanations. The average score for these explanations for coherency, interestingness level, and relevance was 6.33, 5.5, 5.33 respectively across all three datasets. 
From the automatically generated explanations, those generated by \sys\ were clearly preferred. The average score of the explanations generated for the Spotify dataset across the different measures is 5.1, whereas the average score of  \io, \seedb, and \rath\ for this dataset is 3.8, 3, 2.8 respectively. 
For the Credit Card Customers dataset, the average score for \sys\ across the different measures is 5.6, compared to an average score of 4.4, 3.3, 2.9 for \io, \seedb, and \rath, respectively. 
Finally, for the Products and Sales dataset, the average score of the explanations generated by \sys\ across the different measures is 5.3, whereas the average score of \io\ and \seedb\ explanations for this dataset are 3.2, 3.8, respectively. \rath\ timed out for every query over this dataset (after more than 3 hours of computation), and thus its explanations were excluded. 
It should be noted that \sys\ generates hybrid explanations that include both visualization and text, making them easily understandable. 
Explanations generated by other baselines may be harder to interpret since they only contain visualizations or text. 
This may explain the users' preference for these explanations over the explanations generated by some of the baselines.
Interestingly, the scores obtained by \sys\ were very close to these of \expert\ on the Products and Sales Notebook. \sys\ showed such high scores for this notebook due to the join operation; \expert\ did not explain this join while \sys\ noticed a change in the distribution and pointed that out to the user. \sys\ generated more insightful explanations (an average score of 5.4 compared to an average of 5.1 for \expert) and slightly more relevant to the task (average of 5.1 compared to 5 for \expert); in contrast, the \expert\ baseline got higher coherency scores with an average of 6.3 compared to an average of 5.4 for \sys.

\paratitle{\reva{Comparison to unassisted EDA}}
\reva{
We have compared 
standard unassisted EDA with EDA assisted by \sys/\sysopt.
The chosen datasets for the study were Spotify and Credit Card Customers. 
The task for the Spotify dataset was to find which songs are more popular and what makes them popular and the task for the Credit Card Customers dataset was to find out why people leave the service and how can we anticipate it. 
The study included 8 participants who were asked to identify as many insights as possible that are related to the task, when presented with an empty notebook. Then, an expert was asked to denote for each user-generated insight, whether this insight is correct and directly related to the task or not. For example, for the Spotify dataset, "acoustic songs (with acousticness > 0.5) are usually less popular" is a correct insight while "songs from the 90s are louder than other decades" is an unrelated insight. We counted the number of insights gathered by the participants over 10 minutes. 
The average number of insights for the two datasets is shown in Figure \ref{fig:user_study_insights}. 
The average numbers were 2.5 (9.5) with \sysopt\ and 1 (2.5) without it for the Credit Card (Spotify) dataset. 
The results clearly show the benefit of using \sys\ and \sysopt\ and indicate that \sys\ is able to assist users in gaining insights about the EDA task. 
}

\paratitle{Comparison to augmented baselines}
\shep{
We have further added textual explanations to the \seedb\ and \rath\ baselines (in addition to their `organic' visualizations) and performed an additional user study in which the output of all baselines contain both a visualization and a caption. 
In the study, four participants were asked to analyze the Credit Card Customers dataset and its notebook from the first user study (containing the five relevant queries that can be found in \cref{tbl:queries}).
Since the quality of automatic captioning methods may vary, we have asked an expert to manually devise a textual description to the visualizations generated by \seedb\ and \rath{} the visualizations and captions can be found in our code repository \cite{repo}).  
The results are shown in Figure \ref{fig:NL_baseline_userstudy} and indicate that even with experts-generated textual explanations for the baselines, \sys\ is able to generate explanations that are significantly more coherent, insightful, and useful. In particular, the scores were $5.52$ for \sys, $3.17$ for \seedb\ augmented with textual explanations, and $3.42$ for \rath\ augmented with textual explanations. 
}

\subsection{Simulated Experiments}\label{sec:automatic-exp}
\common{We first measure the accuracy of the explanations generated by \sysopt. After establishing that \sysopt\ is indistinguishable from \sys\ in terms of accuracy, we measure its runtimes compared to the baselines when varying the parameters of the database and further measure the trade-off between the number of sets-of-rows and accuracy.}
In the following experiments,
the number of sets-of-rows was set to either $5$ or $10$ (\sys\ tries to divide the values into both $5$ and $10$ sets-of-rows and computes the skylines for all options). For each measured point, we run \sys\ and/or \sysopt\ (depending on the experiment) three times for each one of the relevant queries.

\paratitle{Accuracy of \sysopt}
In this set of experiments, we have averaged over the join/filter queries 1-10 and group-by queries 16-25 from \cref{tbl:gb_queries}, for the Products and Spotify datasets.
Figure \ref{fig:precision_at_k} depicts the precision@k \cite{schutze2008introduction} of \sysopt\ w.r.t. the output of \sys{} ($k$ was set to $3$ since in most cases the number of explanations in the skyline set was $\leq 3$), used as the ground truth. The measurements were performed over the Spotify and Products datasets, as the Credit Card dataset is too small. 
The average precision over the different queries is very high, reaching over 93\%, 96\%, 97\%, and 99\% for sample sizes of 5K, 10K, 20K, and 50K, respectively. This suggests that a relatively small sample can accurately predict the set of skyline explanations. 
Figure \ref{fig:kendall_tau} shows the Kandall-Tau distance \cite{kendall1948rank} between the ground truth results and the results obtained by \sysopt,
when considering all explanations in the skyline. The distance gradually decreases and falls from $74.8$ at sample size $50$ to $10.8$ for sample size 50K. In particular, for a 5K sample the distance is $21.6$ -- lower than the mean value of $33.1$ among all sample sizes. 
Figure \ref{fig:ndcg} shows the nDCG score \cite{JarvelinK02} of \sysopt. 
The starting point of the score is already high (92.6\%), and it further increases gradually with the sample size where the most significant increase is observable as we move from sample sizes of $50$ to $200$ (92\% to 97\%, respectively). In particular, for a sample size of 5K, the nDCG score is 99.8\%. 

\revc{
We have further measured the change in accuracy of \sysopt\ with a sample size of 5K for a varying number of rows for the Products and Sales dataset, which is the largest dataset out of the three. 
Our results are shown in Figure \ref{fig:accuracy_rows}. All three subfigures show that the accuracy remains high for all sets of rows w.r.t. all three metrics. In particular, for 3M rows, the precision@k was $0.942$, the Kendall-Tau distance was $8.1$, and the nDCG value was $0.9985$. Kendall-Tau has minor fluctuations since the different columns have very close interestingness scores, and Kendall-Tau is sensitive to this, as a slight change in interestingness can lead to a change in the ranking of the explanations.
}
{\em In light of these results, in our scalability experiments, we have examined \sysopt\ with a fixed sample size of 5K rows. } 

\shep{
Last,  while an inaccurate interestingness calculation by \sysopt\ may result in sub-optimal explanations, recall from Section~\ref{sec:user-study} that \sysopt\ produced the exact same output as \sys.
In general, there are no guarantees that the skyline will be the same for \sys\ and \sysopt. To explain the reason for the lack of change in the skyline in Section~\ref{sec:user-study}, recall that the skyline set of explanations is determined via the following two steps: (1) computing interestingness and for the top-k most interesting columns (2) computing the contribution for all sets of rows. 
Note that Figure \ref{fig:accuracy_samples} measures only the accuracy for step (1), which is where we see a small loss by the sampling. This loss was not significant enough to affect the final skyline set in Section~\ref{sec:user-study}.
}

\begin{figure}[t]
    \centering
    \includegraphics[width=2in]{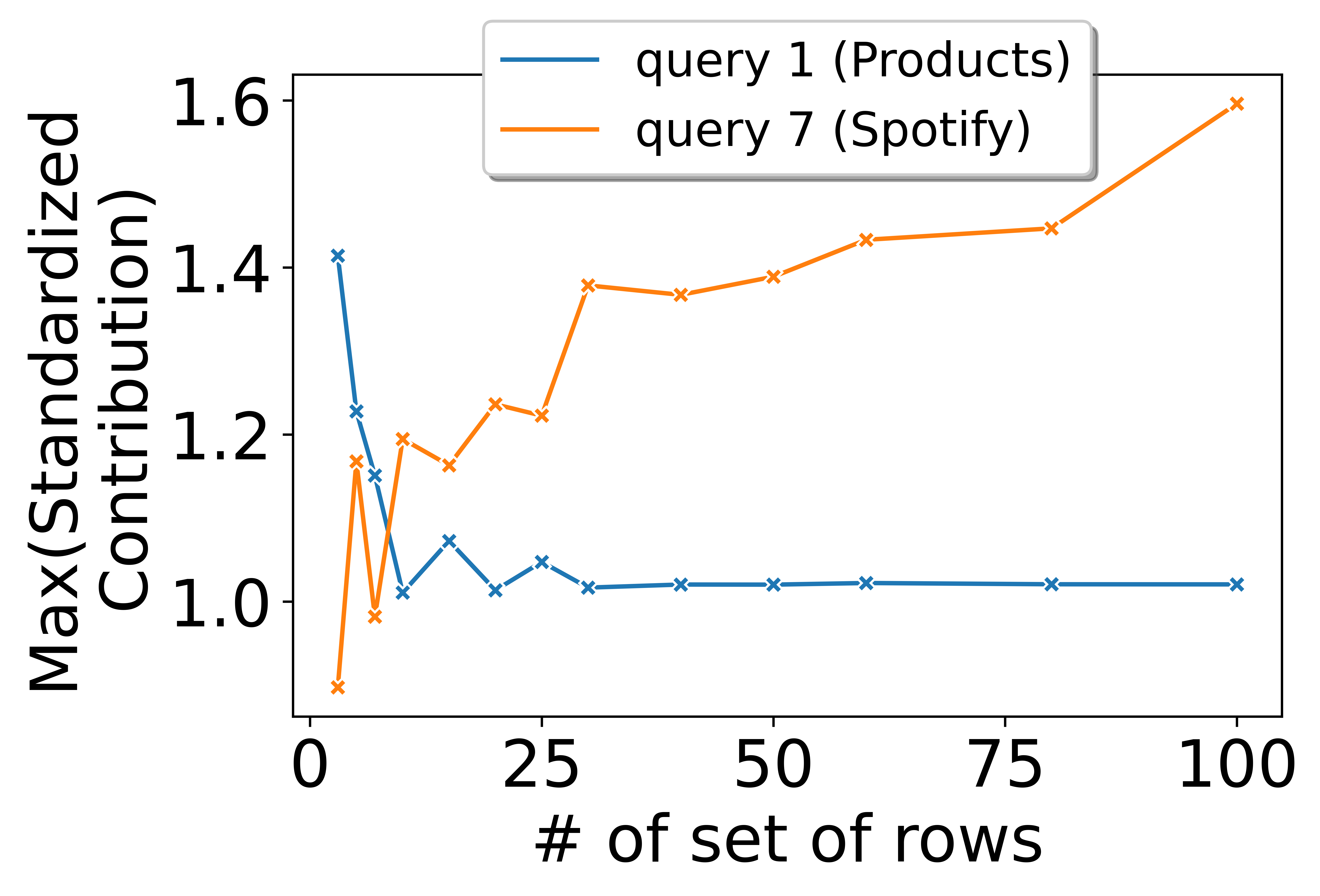}
    \caption{Contribution score (see Section \ref{sec:binning}) for varying number of sets-of-rows for query 1 over Products and Sales, and query 7 over Spotify (see \cref{tbl:queries} for the queries)}
    \label{fig:bin_size}
\end{figure}

\paratitle{Runtime analysis for varying column number}
In this set of experiments, we have examined the runtime as a function of the number of columns in the dataset. The experiment was run over all three datasets, where we averaged over the join/filter queries the 1-15 in \cref{tbl:queries}. 
We have set the number of rows to its maximum size for each dataset and have gradually increased the number of columns. 
We always included two attributes - the attribute that the query needs (e.g., $X$) and the attribute with the highest interestingness score (e.g., $Y$). Then, we perform a random permutation on the attributes and add the columns in a fixed order (if the permutation is $\attr_1,\attr_2,\ldots, \attr_m$, we first check $X,Y,\attr_1$, and then $X,Y,\attr_1,\attr_2$ and so on). 
\revc{
Figure \ref{fig:runtime_columns} depicts the execution time \revc{of \sysopt\ compared to the baselines \seedb\ and \rath}. 
Figure \ref{fig:runtime_columns_bank} shows the runtime for an increasing number of columns for the Credit Card dataset. For 20 columns, the runtime of \sysopt\ was $0.23$s, whereas for \seedb\ and \rath\ it was $0.54$s and $0.52$s, respectively. 
In Figure \ref{fig:runtime_columns_spotify}, at the maximum number of columns for the Spotify dataset (20 columns), 
the runtime of \sysopt\ was 2.27s, while the runtime for \seedb\ and \rath\ was 0.75s, 2.9s, respectively. 
\seedb\ performs better for this specific dataset because it contains mostly numeric attributes, and \seedb\ counts on both categorical values for grouping and numeric attributes for aggregations. The lack of categorical attributes reduces the number of possible views and reduces the runtime.
For the Products dataset (Figure \ref{fig:runtime_columns_products}) with 33 columns, the runtime was 
13.3s for \sysopt\ and 25.1 for \seedb. 
\rath\ was not able to run on 3M rows due to high memory usage (taking more than 17GB, resulting in an `out of memory' error) and long processing times (average of 50\% CPU usage for more than 10 minutes), therefore omitted from the graph. 
These results suggest that \sysopt\ with a sample size of 5K performs better than \seedb\ and \rath\ on datasets with a moderate to large schema size when interactive performance is desired. 
}

\paratitle{Execution time analysis for varying row number}
Here, we examine the execution time as a function of the number of rows for the three datasets. 
In this set of experiments, we have averaged over queries 1-30 from \cref{sec:queries}, separating between group-by queries and filter/join queries.
Again, the sample size of \sysopt\ was set to 5K. 
\revc{Here, for Figure \ref{fig:runtime_rows_products}, we sampled additional rows to increase the size of the view to a maximum of 10M.}

\revc{
Figure~\ref{fig:runtime_rows_bank} shows that \sysopt\ performs better than \seedb\ and \rath\ in most cases as the data size increases with a relatively small increase in execution time as the number of rows grows. For 10K rows, the execution time is 0.23s,
as opposed to an execution time of 0.63s, 0.81s for \seedb\ and \rath, respectively. 
In Figure \ref{fig:runtime_rows_spotify}, for $174,389$ rows, the execution time of 
\sys, \seedb, and \rath\ were 1.81s, 0.7s, 6.4s, respectively. 
\seedb\ slightly outperforms \sysopt\ for this dataset for the same reason detailed in the previous paragraph regarding Figure \ref{fig:runtime_columns_spotify}.
In Figure \ref{fig:runtime_rows_products}, for $10$M rows, 
the execution times of \sysopt\ and \seedb\ were  62.4s, 154.9s, respectively. 
\rath\ was not able to run on these data sizes (as in the explanation for Figure \ref{fig:runtime_columns}, due to high memory usage and large processing times), so it does not appear in the figure. 
This experiment shows that \sysopt\ performs better or in a comparative manner with the examined baselines in terms of runtime for large data sizes. 
}

\paratitle{Accuracy for varying sets-of-rows sizes}
Figure \ref{fig:bin_size} shows the contribution score w.r.t. the numbers of sets-of-rows (see Section \ref{sec:skyline}) for queries 3 and 7 from \cref{tbl:queries} for the Products and Sales and Spotify datasets, respectively. We have chosen two specific queries to ensure that the column will remain constant and that only the number of sets-of-rows will vary. 
There is no clear trend in the results: the optimal number of sets-of-rows w.r.t. contribution depends on the query and the values of the chosen attribute. 
We note that a partition to large number of sets-of-rows resulted in an overloaded and less clear visualization that includes many ticks on the X-axis. Thus, for understandable explanations, choosing a lower number of sets-of-rows appears to be a better strategy (in the user study we set it to 5 or 10). 

%% file: conclusions.tex
\section{Conclusions and Future Work}\label{sec:conclusions}
We have presented \sys, a system that provides coherent explanations for data exploration steps.
The resulted explanations assist users in understanding what is interesting in the results of their exploration steps and gather actionable insights. Our explanations are based on a novel approach, that holistically examines an exploration step and calculates the contribution of sets-of-rows from the \textit{input dataframe} to the interestingness score of a column in the \textit{output dataframe}.
Those sets-of-rows that significantly contribute to highly interesting attributes are transformed to coherent explanations in the form of visualization with a natural language caption.
We have performed an extensive evaluation of our approach that shows its usefulness over real-world datasets and tasks.


There are several interesting directions for future work. One extension of our approach is to allow for different ways of measuring the contribution of row sets, such as causal responsibility.  
Another intriguing direction is to develop new measures that capture additional interestingness facets, as well as to study dedicated optimization techniques for interestingness calculations.
We also plan to design more complex explanations that cover multiple columns and present them in a coherent manner. Finally, we plan to develop a wrapper for Pandas that will allow users to type a simple command in order to explain exploratory operations in one line of code. 

%% file: appendix.tex
\appendix

\section{Queries for the experiments}\label{sec:queries}

\begin{table}[h]
    \centering \footnotesize
    \caption{Join and filter queries used in the experiments (used with the exceptionality measure from Section \ref{sec:interestingness})}\label{tbl:queries}
    \begin{tabularx}{\linewidth}{| c | c | X | c | c | c |}
        \hline {\bf Num.} & {\bf Dataset and Type} & {\bf Query} \\
        \hline 1 & Products (J) & SELECT * FROM products INNER JOIN sales ON products.item$=$sales.item; \\
        
        \hline 2 & Products (J) & SELECT * FROM counties INNER JOIN sales ON counties.county$=$sales.county; \\
        
        \hline 3 & Products (J) & SELECT * FROM counties INNER SELECT * FROM stores INNER JOIN sales ON stores.store=sales.store;\\
        
        \hline 4 & Products (F) & SELECT * FROM products\_sales WHERE sales\_liter\_size $\leq$ 500; \\
        
        \hline 5 & Products (F) & SELECT * FROM products\_sales WHERE sales\_pack $==$ 12; \\
        
        
        
        
        
        
        \hline
        \hline 6 & Spotify (F) & SELECT * FROM spotify WHERE popularity $>$ 65; \\
        
        \hline 7 & Spotify (F) & SELECT * FROM spotify WHERE year $>$ 1990; \\
        
        \hline 8 &  Spotify (F) & SELECT * FROM spotify WHERE loudness $>$ -12;\\
        
        \hline 9 & Spotify (F) & SELECT * FROM spotify WHERE duration\_minutes $<$ 3;\\
        
        \hline 10 & Spotify (F) & SELECT * FROM spotify WHERE tempo $>$ 100;\\
       
       
        
        
        
       
        \hline
        \hline 11 & Bank (F) & SELECT * FROM Bank WHERE Attrition\_Flag != "Existing Customer"; \\
        
        \hline 12 & Bank (F) & SELECT * FROM [SELECT * FROM Bank WHERE Attrition\_Flag != 'Existing Customer'] WHERE Total\_Count\_Change\_Q4\_vs\_Q1 > 0.75; \\
        
        \hline 13 & Bank (F) & SELECT * FROM Bank WHERE Months\_Inactive\_Count\_Last\_Year > 2; \\
        
        \hline 14 & Bank (F) & SELECT * FROM Bank WHERE Customer\_Age < 30; \\
        
        \hline 15 & Bank (F) & SELECT * FROM Bank WHERE Income\_Category $==$ ``Less than \$40K''; \\
        
        
        
        
        
        \hline
    \end{tabularx}
\end{table}

\begin{table}[b]
    \centering \footnotesize
    \caption{Group-by queries used in the experiments (used with the diversity measure from Section \ref{sec:interestingness})}\label{tbl:gb_queries}
    \begin{tabularx}{\linewidth}{| c | c | X | c | c | c |}
        \hline {\bf Num.} & {\bf Dataset and Type} & {\bf Query} \\
        \hline 16 & Products (GB) & SELECT count(item) FROM products\_sales GROUP BY sales\_vendor;\\
        
        \hline 17 & Products (GB) & SELECT count(item) FROM products\_sales GROUP BY sales\_county, sales\_category\_name;\\
        
        \hline 18 & Products (GB) & SELECT count(item) FROM products\_sales GROUP BY products\_sales\_pack;\\
        
        \hline 19 & Products (GB) & SELECT mean(sales\_total), mean(sales\_pack) FROM products\_sales GROUP BY sales\_bottle\_quantity;\\
        
        \hline 20 & Products (GB) & SELECT mean(products\_bottle\_size) FROM products\_sales GROUP BY products\_pack, products\_inner\_pack;\\
        
        \hline
        \hline 21 & Spotify (GB) & SELECT mean(popularity), max(popularity), min(popularity) FROM spotify GROUP BY year;\\
       
        \hline 22 & Spotify (GB) & SELECT mean(danceability), max(danceability), mean(instrumentalness), max(instrumentalness), mean(liveness) FROM spotify GROUP BY year; \\
        
        \hline 23 & Spotify (GB) & SELECT mean(danceability), mean(popularity) FROM spotify GROUP BY key; \\
        
        \hline 24 & Spotify (GB) & SELECT max(duration\_minutes), mean(duration\_minutes) FROM spotify GROUP BY decade; \\
        
        \hline 25 & Spotify (GB) & SELECT mean(loudness), mean(liveness), mean(tempo) FROM spotify GROUP BY mode, key; \\
       
        \hline
        \hline 26 & Bank (GB) & SELECT mean(Credit\_Used), mean(Total\_Transitions\_Amount) FROM Bank GROUP BY Marital\_Status, Income\_Category; \\
        
        \hline 27 & Bank (GB) & SELECT count FROM Bank GROUP BY Marital\_Status, Gender, Education\_Level; \\
        
        \hline 28 & Bank (GB) & SELECT mean(Credit\_Used), mean(Total\_Transitions\_Amount) FROM Bank GROUP BY Marital\_Status; \\
        
        \hline 29 & Bank (GB) & SELECT mean(Customer\_Age) FROM Bank GROUP BY Gender, Income\_Category; \\
        
        \hline 30 & Bank (GB) & SELECT count FROM Bank GROUP BY Registered\_Products\_Count, Attrition\_Flag; \\
        \hline
    \end{tabularx}
\end{table}

%% file: main.bbl

\begin{thebibliography}{81}


\ifx \showCODEN    \undefined \def \showCODEN     #1{\unskip}     \fi
\ifx \showDOI      \undefined \def \showDOI       #1{#1}\fi
\ifx \showISBNx    \undefined \def \showISBNx     #1{\unskip}     \fi
\ifx \showISBNxiii \undefined \def \showISBNxiii  #1{\unskip}     \fi
\ifx \showISSN     \undefined \def \showISSN      #1{\unskip}     \fi
\ifx \showLCCN     \undefined \def \showLCCN      #1{\unskip}     \fi
\ifx \shownote     \undefined \def \shownote      #1{#1}          \fi
\ifx \showarticletitle \undefined \def \showarticletitle #1{#1}   \fi
\ifx \showURL      \undefined \def \showURL       {\relax}        \fi
\providecommand\bibfield[2]{#2}
\providecommand\bibinfo[2]{#2}
\providecommand\natexlab[1]{#1}
\providecommand\showeprint[2][]{arXiv:#2}

\bibitem[\protect\citeauthoryear{Agarwal, Barman, Gunopulos, Young, Korn, and
  Srivastava}{Agarwal et~al\mbox{.}}{2007}]%
        {agarwal2007efficient}
\bibfield{author}{\bibinfo{person}{Deepak Agarwal}, \bibinfo{person}{Dhiman
  Barman}, \bibinfo{person}{Dimitrios Gunopulos}, \bibinfo{person}{Neal~E
  Young}, \bibinfo{person}{Flip Korn}, {and} \bibinfo{person}{Divesh
  Srivastava}.} \bibinfo{year}{2007}\natexlab{}.
\newblock \showarticletitle{Efficient and effective explanation of change in
  hierarchical summaries}. In \bibinfo{booktitle}{\emph{Proceedings of the 13th
  ACM SIGKDD international conference on Knowledge discovery and data mining}}.
  \bibinfo{pages}{6--15}.
\newblock


\bibitem[\protect\citeauthoryear{Amsterdamer, Deutch, and Tannen}{Amsterdamer
  et~al\mbox{.}}{2011}]%
        {amsterdamer2011provenance}
\bibfield{author}{\bibinfo{person}{Yael Amsterdamer}, \bibinfo{person}{Daniel
  Deutch}, {and} \bibinfo{person}{Val Tannen}.}
  \bibinfo{year}{2011}\natexlab{}.
\newblock \showarticletitle{Provenance for aggregate queries}. In
  \bibinfo{booktitle}{\emph{Proceedings of the thirtieth ACM
  SIGMOD-SIGACT-SIGART symposium on Principles of database systems}}.
  \bibinfo{pages}{153--164}.
\newblock


\bibitem[\protect\citeauthoryear{Asudeh, Jagadish, Wu, and Yu}{Asudeh
  et~al\mbox{.}}{2020}]%
        {asudeh2020detecting}
\bibfield{author}{\bibinfo{person}{Abolfazl Asudeh},
  \bibinfo{person}{Hosagrahar~Visvesvaraya Jagadish}, \bibinfo{person}{You Wu},
  {and} \bibinfo{person}{Cong Yu}.} \bibinfo{year}{2020}\natexlab{}.
\newblock \showarticletitle{On detecting cherry-picked trendlines}.
\newblock \bibinfo{journal}{\emph{Proceedings of the VLDB Endowment}}
  \bibinfo{volume}{13}, \bibinfo{number}{6} (\bibinfo{year}{2020}),
  \bibinfo{pages}{939--952}.
\newblock


\bibitem[\protect\citeauthoryear{Bao, Zeng, Jagadish, and Ling}{Bao
  et~al\mbox{.}}{2015}]%
        {bao2015exploratory}
\bibfield{author}{\bibinfo{person}{Zhifeng Bao}, \bibinfo{person}{Yong Zeng},
  \bibinfo{person}{HV Jagadish}, {and} \bibinfo{person}{Tok~Wang Ling}.}
  \bibinfo{year}{2015}\natexlab{}.
\newblock \showarticletitle{Exploratory keyword search with interactive input}.
  In \bibinfo{booktitle}{\emph{Proceedings of the 2015 ACM SIGMOD International
  Conference on Management of Data}}. \bibinfo{pages}{871--876}.
\newblock


\bibitem[\protect\citeauthoryear{Bar~El, Milo, and Somech}{Bar~El
  et~al\mbox{.}}{2020}]%
        {bar2020automatically}
\bibfield{author}{\bibinfo{person}{Ori Bar~El}, \bibinfo{person}{Tova Milo},
  {and} \bibinfo{person}{Amit Somech}.} \bibinfo{year}{2020}\natexlab{}.
\newblock \showarticletitle{Automatically generating data exploration sessions
  using deep reinforcement learning}. In \bibinfo{booktitle}{\emph{SIGMOD}}.
  \bibinfo{pages}{1527--1537}.
\newblock


\bibitem[\protect\citeauthoryear{Barowy, Gochev, and Berger}{Barowy
  et~al\mbox{.}}{2014}]%
        {barowy2014checkcell}
\bibfield{author}{\bibinfo{person}{Daniel~W Barowy}, \bibinfo{person}{Dimitar
  Gochev}, {and} \bibinfo{person}{Emery~D Berger}.}
  \bibinfo{year}{2014}\natexlab{}.
\newblock \showarticletitle{Checkcell: Data debugging for spreadsheets}.
\newblock \bibinfo{journal}{\emph{ACM SIGPLAN Notices}} \bibinfo{volume}{49},
  \bibinfo{number}{10} (\bibinfo{year}{2014}), \bibinfo{pages}{507--523}.
\newblock


\bibitem[\protect\citeauthoryear{Bedeian and Mossholder}{Bedeian and
  Mossholder}{2000}]%
        {bedeian2000use}
\bibfield{author}{\bibinfo{person}{Arthur~G Bedeian} {and}
  \bibinfo{person}{Kevin~W Mossholder}.} \bibinfo{year}{2000}\natexlab{}.
\newblock \showarticletitle{On the use of the coefficient of variation as a
  measure of diversity}.
\newblock \bibinfo{journal}{\emph{Organizational Research Methods}}
  \bibinfo{volume}{3}, \bibinfo{number}{3} (\bibinfo{year}{2000}),
  \bibinfo{pages}{285--297}.
\newblock


\bibitem[\protect\citeauthoryear{Behar and Cohen}{Behar and Cohen}{2020a}]%
        {behar2020optimal_cikm}
\bibfield{author}{\bibinfo{person}{Rachel Behar} {and} \bibinfo{person}{Sara
  Cohen}.} \bibinfo{year}{2020}\natexlab{a}.
\newblock \showarticletitle{Optimal End-Biased Histograms for Hierarchical
  Data}. In \bibinfo{booktitle}{\emph{Proceedings of the 29th ACM International
  Conference on Information \& Knowledge Management}}.
  \bibinfo{pages}{3261--3264}.
\newblock


\bibitem[\protect\citeauthoryear{Behar and Cohen}{Behar and Cohen}{2020b}]%
        {behar2020optimale_edbt}
\bibfield{author}{\bibinfo{person}{Rachel Behar} {and} \bibinfo{person}{Sara
  Cohen}.} \bibinfo{year}{2020}\natexlab{b}.
\newblock \showarticletitle{Optimal Histograms with Outliers.}. In
  \bibinfo{booktitle}{\emph{Extending database technology (EDBT)}}.
  \bibinfo{pages}{181--192}.
\newblock


\bibitem[\protect\citeauthoryear{Bespinyowong, Chen, Jagadish, and
  Ma}{Bespinyowong et~al\mbox{.}}{2016}]%
        {bespinyowong2016exrank}
\bibfield{author}{\bibinfo{person}{Ramon Bespinyowong}, \bibinfo{person}{Wei
  Chen}, \bibinfo{person}{HV Jagadish}, {and} \bibinfo{person}{Yuxin Ma}.}
  \bibinfo{year}{2016}\natexlab{}.
\newblock \showarticletitle{ExRank: An exploratory ranking interface}.
\newblock \bibinfo{journal}{\emph{PVLBD}} \bibinfo{volume}{9},
  \bibinfo{number}{13} (\bibinfo{year}{2016}), \bibinfo{pages}{1529--1532}.
\newblock


\bibitem[\protect\citeauthoryear{Bidoit, Herschel, and Tzompanaki}{Bidoit
  et~al\mbox{.}}{2015}]%
        {bidoit2015efficient}
\bibfield{author}{\bibinfo{person}{Nicole Bidoit}, \bibinfo{person}{Melanie
  Herschel}, {and} \bibinfo{person}{Aikaterini Tzompanaki}.}
  \bibinfo{year}{2015}\natexlab{}.
\newblock \showarticletitle{Efficient computation of polynomial explanations of
  why-not questions}. In \bibinfo{booktitle}{\emph{Proceedings of the 24th ACM
  International on Conference on Information and Knowledge Management}}.
  \bibinfo{pages}{713--722}.
\newblock


\bibitem[\protect\citeauthoryear{Bidoit, Herschel, and Tzompanaki}{Bidoit
  et~al\mbox{.}}{2014}]%
        {bidoit2014query}
\bibfield{author}{\bibinfo{person}{Nicole Bidoit}, \bibinfo{person}{Melanie
  Herschel}, {and} \bibinfo{person}{Katerina Tzompanaki}.}
  \bibinfo{year}{2014}\natexlab{}.
\newblock \showarticletitle{Query-based why-not provenance with nedexplain}. In
  \bibinfo{booktitle}{\emph{Extending database technology (EDBT)}}.
\newblock


\bibitem[\protect\citeauthoryear{Borzsony, Kossmann, and Stocker}{Borzsony
  et~al\mbox{.}}{2001}]%
        {borzsony2001skyline}
\bibfield{author}{\bibinfo{person}{Stephan Borzsony}, \bibinfo{person}{Donald
  Kossmann}, {and} \bibinfo{person}{Konrad Stocker}.}
  \bibinfo{year}{2001}\natexlab{}.
\newblock \showarticletitle{The skyline operator}. In
  \bibinfo{booktitle}{\emph{Proceedings 17th international conference on data
  engineering}}. IEEE, \bibinfo{pages}{421--430}.
\newblock


\bibitem[\protect\citeauthoryear{Brown}{Brown}{2011}]%
        {brown2011measures}
\bibfield{author}{\bibinfo{person}{Stan Brown}.}
  \bibinfo{year}{2011}\natexlab{}.
\newblock \bibinfo{title}{Measures of shape: Skewness and kurtosis}.
\newblock
\newblock


\bibitem[\protect\citeauthoryear{Buneman, Khanna, and Tan}{Buneman
  et~al\mbox{.}}{2001}]%
        {why}
\bibfield{author}{\bibinfo{person}{P. Buneman}, \bibinfo{person}{S. Khanna},
  {and} \bibinfo{person}{W.C. Tan}.} \bibinfo{year}{2001}\natexlab{}.
\newblock \showarticletitle{Why and Where: A Characterization of Data
  Provenance}. In \bibinfo{booktitle}{\emph{ICDT}}. \bibinfo{pages}{316--330}.
\newblock


\bibitem[\protect\citeauthoryear{Chandola and Kumar}{Chandola and
  Kumar}{2007}]%
        {chandola2007summarization}
\bibfield{author}{\bibinfo{person}{Varun Chandola} {and} \bibinfo{person}{Vipin
  Kumar}.} \bibinfo{year}{2007}\natexlab{}.
\newblock \showarticletitle{Summarization--compressing data into an informative
  representation}.
\newblock \bibinfo{journal}{\emph{Knowledge and Information Systems}}
  \bibinfo{volume}{12}, \bibinfo{number}{3} (\bibinfo{year}{2007}),
  \bibinfo{pages}{355--378}.
\newblock


\bibitem[\protect\citeauthoryear{Chapman and Jagadish}{Chapman and
  Jagadish}{2009}]%
        {chapman2009not}
\bibfield{author}{\bibinfo{person}{Adriane Chapman} {and} \bibinfo{person}{HV
  Jagadish}.} \bibinfo{year}{2009}\natexlab{}.
\newblock \showarticletitle{Why not?}. In \bibinfo{booktitle}{\emph{Proceedings
  of the 2009 ACM SIGMOD International Conference on Management of data}}.
  \bibinfo{pages}{523--534}.
\newblock


\bibitem[\protect\citeauthoryear{Cui, Zhang, Wang, Huang, Chen, Fang, Zhang,
  Lou, and Zhang}{Cui et~al\mbox{.}}{2019}]%
        {cui2019text}
\bibfield{author}{\bibinfo{person}{Weiwei Cui}, \bibinfo{person}{Xiaoyu Zhang},
  \bibinfo{person}{Yun Wang}, \bibinfo{person}{He Huang}, \bibinfo{person}{Bei
  Chen}, \bibinfo{person}{Lei Fang}, \bibinfo{person}{Haidong Zhang},
  \bibinfo{person}{Jian-Guan Lou}, {and} \bibinfo{person}{Dongmei Zhang}.}
  \bibinfo{year}{2019}\natexlab{}.
\newblock \showarticletitle{Text-to-viz: Automatic generation of infographics
  from proportion-related natural language statements}.
\newblock \bibinfo{journal}{\emph{IEEE transactions on visualization and
  computer graphics}} \bibinfo{volume}{26}, \bibinfo{number}{1}
  (\bibinfo{year}{2019}), \bibinfo{pages}{906--916}.
\newblock


\bibitem[\protect\citeauthoryear{Dataset}{Dataset}{2021a}]%
        {creditdataset}
\bibfield{author}{\bibinfo{person}{Credit Card~Customers Dataset}.}
  \bibinfo{year}{2021}\natexlab{a}.
\newblock
  \bibinfo{howpublished}{\url{https://www.kaggle.com/sakshigoyal7/credit-card-customers/tasks?taskId=2729}}.
\newblock


\bibitem[\protect\citeauthoryear{Dataset}{Dataset}{2021b}]%
        {spotifydataset}
\bibfield{author}{\bibinfo{person}{Spotify Dataset}.}
  \bibinfo{year}{2021}\natexlab{b}.
\newblock
  \bibinfo{howpublished}{\url{https://www.kaggle.com/mrmorj/dataset-of-songs-in-spotify}}.
\newblock


\bibitem[\protect\citeauthoryear{De~Bie}{De~Bie}{2013}]%
        {de2013subjective}
\bibfield{author}{\bibinfo{person}{Tijl De~Bie}.}
  \bibinfo{year}{2013}\natexlab{}.
\newblock \showarticletitle{Subjective interestingness in exploratory data
  mining}.
\newblock In \bibinfo{booktitle}{\emph{Advances in Intelligent Data Analysis
  XII}}. \bibinfo{publisher}{Springer}, \bibinfo{pages}{19--31}.
\newblock


\bibitem[\protect\citeauthoryear{Deutch, Frost, and Gilad}{Deutch
  et~al\mbox{.}}{2017}]%
        {DeutchFG17}
\bibfield{author}{\bibinfo{person}{Daniel Deutch}, \bibinfo{person}{Nave
  Frost}, {and} \bibinfo{person}{Amir Gilad}.} \bibinfo{year}{2017}\natexlab{}.
\newblock \showarticletitle{Provenance for Natural Language Queries}.
\newblock \bibinfo{journal}{\emph{{PVLDB}}} \bibinfo{volume}{10},
  \bibinfo{number}{5} (\bibinfo{year}{2017}), \bibinfo{pages}{577--588}.
\newblock


\bibitem[\protect\citeauthoryear{Deutch and Gilad}{Deutch and Gilad}{2016}]%
        {deutch2016qplain}
\bibfield{author}{\bibinfo{person}{Daniel Deutch} {and} \bibinfo{person}{Amir
  Gilad}.} \bibinfo{year}{2016}\natexlab{}.
\newblock \showarticletitle{QPlain: Query by explanation}. In
  \bibinfo{booktitle}{\emph{2016 IEEE 32nd International Conference on Data
  Engineering (ICDE)}}. IEEE, \bibinfo{pages}{1358--1361}.
\newblock


\bibitem[\protect\citeauthoryear{Dimitriadou, Papaemmanouil, and
  Diao}{Dimitriadou et~al\mbox{.}}{2016}]%
        {dimitriadou2016aide}
\bibfield{author}{\bibinfo{person}{Kyriaki Dimitriadou}, \bibinfo{person}{Olga
  Papaemmanouil}, {and} \bibinfo{person}{Yanlei Diao}.}
  \bibinfo{year}{2016}\natexlab{}.
\newblock \showarticletitle{AIDE: An Active Learning-based Approach for
  Interactive Data Exploration}.
\newblock \bibinfo{journal}{\emph{TKDE}} (\bibinfo{year}{2016}).
\newblock


\bibitem[\protect\citeauthoryear{Ding, Han, Xu, Zhang, and Zhang}{Ding
  et~al\mbox{.}}{2019}]%
        {ding2019quickinsights}
\bibfield{author}{\bibinfo{person}{Rui Ding}, \bibinfo{person}{Shi Han},
  \bibinfo{person}{Yong Xu}, \bibinfo{person}{Haidong Zhang}, {and}
  \bibinfo{person}{Dongmei Zhang}.} \bibinfo{year}{2019}\natexlab{}.
\newblock \showarticletitle{Quickinsights: Quick and automatic discovery of
  insights from multi-dimensional data}. In
  \bibinfo{booktitle}{\emph{Proceedings of the 2019 International Conference on
  Management of Data}}. \bibinfo{pages}{317--332}.
\newblock


\bibitem[\protect\citeauthoryear{Dong and Srivastava}{Dong and
  Srivastava}{2013}]%
        {dong2013compact}
\bibfield{author}{\bibinfo{person}{Xin~Luna Dong} {and} \bibinfo{person}{Divesh
  Srivastava}.} \bibinfo{year}{2013}\natexlab{}.
\newblock \showarticletitle{Compact explanation of data fusion decisions}. In
  \bibinfo{booktitle}{\emph{Proceedings of the 22nd international conference on
  World Wide Web}}. \bibinfo{pages}{379--390}.
\newblock


\bibitem[\protect\citeauthoryear{Fariha, Tiwari, Radhakrishna, and
  Gulwani}{Fariha et~al\mbox{.}}{2020}]%
        {fariha2020extune}
\bibfield{author}{\bibinfo{person}{Anna Fariha}, \bibinfo{person}{Ashish
  Tiwari}, \bibinfo{person}{Arjun Radhakrishna}, {and} \bibinfo{person}{Sumit
  Gulwani}.} \bibinfo{year}{2020}\natexlab{}.
\newblock \showarticletitle{Extune: Explaining tuple non-conformance}. In
  \bibinfo{booktitle}{\emph{Proceedings of the 2020 ACM SIGMOD International
  Conference on Management of Data}}. \bibinfo{pages}{2741--2744}.
\newblock


\bibitem[\protect\citeauthoryear{Geng and Hamilton}{Geng and Hamilton}{2006}]%
        {geng2006interestingness}
\bibfield{author}{\bibinfo{person}{Liqiang Geng} {and}
  \bibinfo{person}{Howard~J Hamilton}.} \bibinfo{year}{2006}\natexlab{}.
\newblock \showarticletitle{Interestingness measures for data mining: A
  survey}.
\newblock \bibinfo{journal}{\emph{ACM Computing Surveys (CSUR)}}
  \bibinfo{volume}{38}, \bibinfo{number}{3} (\bibinfo{year}{2006}),
  \bibinfo{pages}{9--es}.
\newblock


\bibitem[\protect\citeauthoryear{Green, Karvounarakis, and Tannen}{Green
  et~al\mbox{.}}{2007}]%
        {GKT-pods07}
\bibfield{author}{\bibinfo{person}{T.J. Green}, \bibinfo{person}{G.
  Karvounarakis}, {and} \bibinfo{person}{V. Tannen}.}
  \bibinfo{year}{2007}\natexlab{}.
\newblock \showarticletitle{Provenance semirings}. In
  \bibinfo{booktitle}{\emph{PODS}}. \bibinfo{pages}{31--40}.
\newblock


\bibitem[\protect\citeauthoryear{Harris, Millman, van~der Walt, Gommers,
  Virtanen, Cournapeau, Wieser, Taylor, Berg, Smith, Kern, Picus, Hoyer, van
  Kerkwijk, Brett, Haldane, del R{\'{i}}o, Wiebe, Peterson,
  G{\'{e}}rard-Marchant, Sheppard, Reddy, Weckesser, Abbasi, Gohlke, and
  Oliphant}{Harris et~al\mbox{.}}{2020}]%
        {harris2020array}
\bibfield{author}{\bibinfo{person}{Charles~R. Harris},
  \bibinfo{person}{K.~Jarrod Millman}, \bibinfo{person}{St{\'{e}}fan~J. van~der
  Walt}, \bibinfo{person}{Ralf Gommers}, \bibinfo{person}{Pauli Virtanen},
  \bibinfo{person}{David Cournapeau}, \bibinfo{person}{Eric Wieser},
  \bibinfo{person}{Julian Taylor}, \bibinfo{person}{Sebastian Berg},
  \bibinfo{person}{Nathaniel~J. Smith}, \bibinfo{person}{Robert Kern},
  \bibinfo{person}{Matti Picus}, \bibinfo{person}{Stephan Hoyer},
  \bibinfo{person}{Marten~H. van Kerkwijk}, \bibinfo{person}{Matthew Brett},
  \bibinfo{person}{Allan Haldane}, \bibinfo{person}{Jaime~Fern{\'{a}}ndez del
  R{\'{i}}o}, \bibinfo{person}{Mark Wiebe}, \bibinfo{person}{Pearu Peterson},
  \bibinfo{person}{Pierre G{\'{e}}rard-Marchant}, \bibinfo{person}{Kevin
  Sheppard}, \bibinfo{person}{Tyler Reddy}, \bibinfo{person}{Warren Weckesser},
  \bibinfo{person}{Hameer Abbasi}, \bibinfo{person}{Christoph Gohlke}, {and}
  \bibinfo{person}{Travis~E. Oliphant}.} \bibinfo{year}{2020}\natexlab{}.
\newblock \showarticletitle{Array programming with {NumPy}}.
\newblock \bibinfo{journal}{\emph{Nature}} \bibinfo{volume}{585},
  \bibinfo{number}{7825} (\bibinfo{year}{2020}), \bibinfo{pages}{357--362}.
\newblock


\bibitem[\protect\citeauthoryear{Hilderman and Hamilton}{Hilderman and
  Hamilton}{2013}]%
        {hilderman2013knowledge}
\bibfield{author}{\bibinfo{person}{Robert~J Hilderman} {and}
  \bibinfo{person}{Howard~J Hamilton}.} \bibinfo{year}{2013}\natexlab{}.
\newblock \bibinfo{booktitle}{\emph{Knowledge discovery and measures of
  interest}}. Vol.~\bibinfo{volume}{638}.
\newblock \bibinfo{publisher}{Springer Science \& Business Media}.
\newblock


\bibitem[\protect\citeauthoryear{Huang, Yan, Lu, Lin, Gao, and Chen}{Huang
  et~al\mbox{.}}{2019}]%
        {huang2019leri}
\bibfield{author}{\bibinfo{person}{Hao Huang}, \bibinfo{person}{Qian Yan},
  \bibinfo{person}{Wei Lu}, \bibinfo{person}{Huaizhong Lin},
  \bibinfo{person}{Yunjun Gao}, {and} \bibinfo{person}{Lei Chen}.}
  \bibinfo{year}{2019}\natexlab{}.
\newblock \showarticletitle{LERI: Local Exploration for Rare-Category
  Identification}.
\newblock \bibinfo{journal}{\emph{IEEE Transactions on Knowledge and Data
  Engineering}} \bibinfo{volume}{32}, \bibinfo{number}{9}
  (\bibinfo{year}{2019}), \bibinfo{pages}{1761--1772}.
\newblock


\bibitem[\protect\citeauthoryear{Hunter}{Hunter}{2007}]%
        {matplotlib}
\bibfield{author}{\bibinfo{person}{J.~D. Hunter}.}
  \bibinfo{year}{2007}\natexlab{}.
\newblock \showarticletitle{Matplotlib: A 2D graphics environment}.
\newblock \bibinfo{journal}{\emph{Computing in Science \& Engineering}}
  \bibinfo{volume}{9}, \bibinfo{number}{3} (\bibinfo{year}{2007}),
  \bibinfo{pages}{90--95}.
\newblock


\bibitem[\protect\citeauthoryear{Ilyas, Markl, Haas, Brown, and
  Aboulnaga}{Ilyas et~al\mbox{.}}{2004}]%
        {ilyas2004cords}
\bibfield{author}{\bibinfo{person}{Ihab~F Ilyas}, \bibinfo{person}{Volker
  Markl}, \bibinfo{person}{Peter Haas}, \bibinfo{person}{Paul Brown}, {and}
  \bibinfo{person}{Ashraf Aboulnaga}.} \bibinfo{year}{2004}\natexlab{}.
\newblock \showarticletitle{CORDS: Automatic discovery of correlations and soft
  functional dependencies}. In \bibinfo{booktitle}{\emph{Proceedings of the
  2004 ACM SIGMOD international conference on Management of data}}.
  \bibinfo{pages}{647--658}.
\newblock


\bibitem[\protect\citeauthoryear{J{\"{a}}rvelin and
  Kek{\"{a}}l{\"{a}}inen}{J{\"{a}}rvelin and Kek{\"{a}}l{\"{a}}inen}{2002}]%
        {JarvelinK02}
\bibfield{author}{\bibinfo{person}{Kalervo J{\"{a}}rvelin} {and}
  \bibinfo{person}{Jaana Kek{\"{a}}l{\"{a}}inen}.}
  \bibinfo{year}{2002}\natexlab{}.
\newblock \showarticletitle{Cumulated gain-based evaluation of {IR}
  techniques}.
\newblock \bibinfo{journal}{\emph{{ACM} Trans. Inf. Syst.}}
  \bibinfo{volume}{20}, \bibinfo{number}{4} (\bibinfo{year}{2002}),
  \bibinfo{pages}{422--446}.
\newblock


\bibitem[\protect\citeauthoryear{Joglekar, Garcia-Molina, and
  Parameswaran}{Joglekar et~al\mbox{.}}{2017}]%
        {joglekar2017interactive}
\bibfield{author}{\bibinfo{person}{Manas Joglekar}, \bibinfo{person}{Hector
  Garcia-Molina}, {and} \bibinfo{person}{Aditya Parameswaran}.}
  \bibinfo{year}{2017}\natexlab{}.
\newblock \showarticletitle{Interactive data exploration with smart
  drill-down}.
\newblock \bibinfo{journal}{\emph{IEEE Transactions on Knowledge and Data
  Engineering}} \bibinfo{volume}{31}, \bibinfo{number}{1}
  (\bibinfo{year}{2017}), \bibinfo{pages}{46--60}.
\newblock


\bibitem[\protect\citeauthoryear{Kendall}{Kendall}{1948}]%
        {kendall1948rank}
\bibfield{author}{\bibinfo{person}{Maurice~George Kendall}.}
  \bibinfo{year}{1948}\natexlab{}.
\newblock \showarticletitle{Rank correlation methods}.
\newblock  (\bibinfo{year}{1948}).
\newblock


\bibitem[\protect\citeauthoryear{Kery, Radensky, Arya, John, and Myers}{Kery
  et~al\mbox{.}}{2018}]%
        {kery2018story}
\bibfield{author}{\bibinfo{person}{Mary~Beth Kery}, \bibinfo{person}{Marissa
  Radensky}, \bibinfo{person}{Mahima Arya}, \bibinfo{person}{Bonnie~E John},
  {and} \bibinfo{person}{Brad~A Myers}.} \bibinfo{year}{2018}\natexlab{}.
\newblock \showarticletitle{The story in the notebook: Exploratory data science
  using a literate programming tool}. In \bibinfo{booktitle}{\emph{CHI}}.
\newblock


\bibitem[\protect\citeauthoryear{Khoussainova, Kwon, Balazinska, and
  Suciu}{Khoussainova et~al\mbox{.}}{2010}]%
        {khoussainova2010snipsuggest}
\bibfield{author}{\bibinfo{person}{Nodira Khoussainova},
  \bibinfo{person}{YongChul Kwon}, \bibinfo{person}{Magdalena Balazinska},
  {and} \bibinfo{person}{Dan Suciu}.} \bibinfo{year}{2010}\natexlab{}.
\newblock \showarticletitle{SnipSuggest: Context-Aware Autocompletion for
  {SQL}}.
\newblock \bibinfo{journal}{\emph{Proc. {VLDB} Endow.}} \bibinfo{volume}{4},
  \bibinfo{number}{1} (\bibinfo{year}{2010}), \bibinfo{pages}{22--33}.
\newblock


\bibitem[\protect\citeauthoryear{Le~Guilly, Petit, Scuturici, and
  Ilyas}{Le~Guilly et~al\mbox{.}}{2019}]%
        {le2019explique}
\bibfield{author}{\bibinfo{person}{Marie Le~Guilly}, \bibinfo{person}{Jean-Marc
  Petit}, \bibinfo{person}{Vasile-Marian Scuturici}, {and}
  \bibinfo{person}{Ihab~F Ilyas}.} \bibinfo{year}{2019}\natexlab{}.
\newblock \showarticletitle{ExplIQuE: Interactive Databases Exploration with
  SQL}. In \bibinfo{booktitle}{\emph{Proceedings of the 28th ACM International
  Conference on Information and Knowledge Management}}.
  \bibinfo{pages}{2877--2880}.
\newblock


\bibitem[\protect\citeauthoryear{Lee, Tang, Agarwal, Boonmark, Chen, Kang,
  Mukhopadhyay, Song, Yong, Hearst, et~al\mbox{.}}{Lee et~al\mbox{.}}{2021}]%
        {lee2021lux}
\bibfield{author}{\bibinfo{person}{Doris Jung-Lin Lee}, \bibinfo{person}{Dixin
  Tang}, \bibinfo{person}{Kunal Agarwal}, \bibinfo{person}{Thyne Boonmark},
  \bibinfo{person}{Caitlyn Chen}, \bibinfo{person}{Jake Kang},
  \bibinfo{person}{Ujjaini Mukhopadhyay}, \bibinfo{person}{Jerry Song},
  \bibinfo{person}{Micah Yong}, \bibinfo{person}{Marti~A Hearst},
  {et~al\mbox{.}}} \bibinfo{year}{2021}\natexlab{}.
\newblock \showarticletitle{Lux: Always-on Visualization Recommendations for
  Exploratory Data Science}.
\newblock \bibinfo{journal}{\emph{arXiv preprint arXiv:2105.00121}}
  (\bibinfo{year}{2021}).
\newblock


\bibitem[\protect\citeauthoryear{Li, Miao, Zeng, Glavic, and Roy}{Li
  et~al\mbox{.}}{2021}]%
        {li2021putting}
\bibfield{author}{\bibinfo{person}{Chenjie Li}, \bibinfo{person}{Zhengjie
  Miao}, \bibinfo{person}{Qitian Zeng}, \bibinfo{person}{Boris Glavic}, {and}
  \bibinfo{person}{Sudeepa Roy}.} \bibinfo{year}{2021}\natexlab{}.
\newblock \showarticletitle{Putting Things into Context: Rich Explanations for
  Query Answers using Join Graphs}. In \bibinfo{booktitle}{\emph{Proceedings of
  the 2021 International Conference on Management of Data}}.
  \bibinfo{pages}{1051--1063}.
\newblock


\bibitem[\protect\citeauthoryear{Liu, Hsu, Mun, and Lee}{Liu
  et~al\mbox{.}}{1999}]%
        {liu1999finding}
\bibfield{author}{\bibinfo{person}{Bing Liu}, \bibinfo{person}{Wynne Hsu},
  \bibinfo{person}{Lai-Fun Mun}, {and} \bibinfo{person}{Hing-Yan Lee}.}
  \bibinfo{year}{1999}\natexlab{}.
\newblock \showarticletitle{Finding interesting patterns using user
  expectations}.
\newblock \bibinfo{journal}{\emph{IEEE Transactions on Knowledge and Data
  Engineering}} \bibinfo{volume}{11}, \bibinfo{number}{6}
  (\bibinfo{year}{1999}), \bibinfo{pages}{817--832}.
\newblock


\bibitem[\protect\citeauthoryear{Liu, Golab, and Ilyas}{Liu
  et~al\mbox{.}}{2015}]%
        {liu2015smas}
\bibfield{author}{\bibinfo{person}{Xiufeng Liu}, \bibinfo{person}{Lukasz
  Golab}, {and} \bibinfo{person}{Ihab~F Ilyas}.}
  \bibinfo{year}{2015}\natexlab{}.
\newblock \showarticletitle{SMAS: A smart meter data analytics system}. In
  \bibinfo{booktitle}{\emph{2015 IEEE 31st International Conference on Data
  Engineering}}. IEEE, \bibinfo{pages}{1476--1479}.
\newblock


\bibitem[\protect\citeauthoryear{Luo, Qin, Tang, and Li}{Luo
  et~al\mbox{.}}{2018}]%
        {luo2018deepeye}
\bibfield{author}{\bibinfo{person}{Yuyu Luo}, \bibinfo{person}{Xuedi Qin},
  \bibinfo{person}{Nan Tang}, {and} \bibinfo{person}{Guoliang Li}.}
  \bibinfo{year}{2018}\natexlab{}.
\newblock \showarticletitle{DeepEye: Towards Automatic Data Visualization}.
  ICDE.
\newblock


\bibitem[\protect\citeauthoryear{McGarry}{McGarry}{2005}]%
        {mcgarry2005survey}
\bibfield{author}{\bibinfo{person}{Ken McGarry}.}
  \bibinfo{year}{2005}\natexlab{}.
\newblock \showarticletitle{A survey of interestingness measures for knowledge
  discovery}.
\newblock \bibinfo{journal}{\emph{The knowledge engineering review}}
  \bibinfo{volume}{20}, \bibinfo{number}{1} (\bibinfo{year}{2005}),
  \bibinfo{pages}{39--61}.
\newblock


\bibitem[\protect\citeauthoryear{Meliou, Gatterbauer, Halpern, Koch, Moore, and
  Suciu}{Meliou et~al\mbox{.}}{2010}]%
        {meliou2010causality}
\bibfield{author}{\bibinfo{person}{Alexandra Meliou}, \bibinfo{person}{Wolfgang
  Gatterbauer}, \bibinfo{person}{Joseph~Y Halpern}, \bibinfo{person}{Christoph
  Koch}, \bibinfo{person}{Katherine~F Moore}, {and} \bibinfo{person}{Dan
  Suciu}.} \bibinfo{year}{2010}\natexlab{}.
\newblock \showarticletitle{Causality in databases}.
\newblock \bibinfo{journal}{\emph{IEEE Data Engineering Bulletin}}
  \bibinfo{volume}{33} (\bibinfo{year}{2010}), \bibinfo{pages}{59--67}.
\newblock


\bibitem[\protect\citeauthoryear{Miao, Zeng, Glavic, and Roy}{Miao
  et~al\mbox{.}}{2019}]%
        {miao2019going}
\bibfield{author}{\bibinfo{person}{Zhengjie Miao}, \bibinfo{person}{Qitian
  Zeng}, \bibinfo{person}{Boris Glavic}, {and} \bibinfo{person}{Sudeepa Roy}.}
  \bibinfo{year}{2019}\natexlab{}.
\newblock \showarticletitle{Going beyond provenance: Explaining query answers
  with pattern-based counterbalances}. In \bibinfo{booktitle}{\emph{Proceedings
  of the 2019 International Conference on Management of Data}}.
  \bibinfo{pages}{485--502}.
\newblock


\bibitem[\protect\citeauthoryear{Milo, Ozeri, and Somech}{Milo
  et~al\mbox{.}}{2019}]%
        {milo2019predicting}
\bibfield{author}{\bibinfo{person}{Tova Milo}, \bibinfo{person}{Chai Ozeri},
  {and} \bibinfo{person}{Amit Somech}.} \bibinfo{year}{2019}\natexlab{}.
\newblock \showarticletitle{Predicting "What is Interesting" by Mining
  Interactive-Data-Analysis Session Logs}. In \bibinfo{booktitle}{\emph{EDBT}}.
  \bibinfo{pages}{456--467}.
\newblock


\bibitem[\protect\citeauthoryear{Milo and Somech}{Milo and Somech}{2018}]%
        {Somech2018REACTKDD}
\bibfield{author}{\bibinfo{person}{Tova Milo} {and} \bibinfo{person}{Amit
  Somech}.} \bibinfo{year}{2018}\natexlab{}.
\newblock \showarticletitle{Next-Step Suggestions for Modern Interactive Data
  Analysis Platforms}. In \bibinfo{booktitle}{\emph{{KDD}}}.
  \bibinfo{publisher}{{ACM}}, \bibinfo{pages}{576--585}.
\newblock


\bibitem[\protect\citeauthoryear{Mirylenka, Cormode, Palpanas, and
  Srivastava}{Mirylenka et~al\mbox{.}}{2015}]%
        {mirylenka2015conditional}
\bibfield{author}{\bibinfo{person}{Katsiaryna Mirylenka},
  \bibinfo{person}{Graham Cormode}, \bibinfo{person}{Themis Palpanas}, {and}
  \bibinfo{person}{Divesh Srivastava}.} \bibinfo{year}{2015}\natexlab{}.
\newblock \showarticletitle{Conditional heavy hitters: detecting interesting
  correlations in data streams}.
\newblock \bibinfo{journal}{\emph{The VLDB Journal}} \bibinfo{volume}{24},
  \bibinfo{number}{3} (\bibinfo{year}{2015}), \bibinfo{pages}{395--414}.
\newblock


\bibitem[\protect\citeauthoryear{Mirylenka, Palpanas, Cormode, and
  Srivastava}{Mirylenka et~al\mbox{.}}{2013}]%
        {mirylenka2013finding}
\bibfield{author}{\bibinfo{person}{Katsiaryna Mirylenka},
  \bibinfo{person}{Themis Palpanas}, \bibinfo{person}{Graham Cormode}, {and}
  \bibinfo{person}{Divesh Srivastava}.} \bibinfo{year}{2013}\natexlab{}.
\newblock \showarticletitle{Finding interesting correlations with conditional
  heavy hitters}. In \bibinfo{booktitle}{\emph{2013 IEEE 29th International
  Conference on Data Engineering (ICDE)}}. IEEE, \bibinfo{pages}{1069--1080}.
\newblock


\bibitem[\protect\citeauthoryear{pandas~development team}{pandas~development
  team}{2020}]%
        {reback2020pandas}
\bibfield{author}{\bibinfo{person}{The pandas~development team}.}
  \bibinfo{year}{2020}\natexlab{}.
\newblock \bibinfo{booktitle}{\emph{pandas-dev/pandas: Pandas}}.
\newblock
\urldef\tempurl%
\url{https://doi.org/10.5281/zenodo.3509134}
\showDOI{\tempurl}


\bibitem[\protect\citeauthoryear{Pearl et~al\mbox{.}}{Pearl
  et~al\mbox{.}}{2009}]%
        {pearl2009causality}
\bibfield{author}{\bibinfo{person}{Judea Pearl} {et~al\mbox{.}}}
  \bibinfo{year}{2009}\natexlab{}.
\newblock \showarticletitle{Causal inference in statistics: An overview}.
\newblock \bibinfo{journal}{\emph{Statistics surveys}}  \bibinfo{volume}{3}
  (\bibinfo{year}{2009}), \bibinfo{pages}{96--146}.
\newblock


\bibitem[\protect\citeauthoryear{Products and dataset}{Products and
  dataset}{2018}]%
        {products}
\bibfield{author}{\bibinfo{person}{Products} {and} \bibinfo{person}{Sales
  dataset}.} \bibinfo{year}{2018}\natexlab{}.
\newblock
  \bibinfo{howpublished}{\url{https://data.world/classrooms/guide-to-data-analysis-with-sql-and-datadotworld}}.
\newblock


\bibitem[\protect\citeauthoryear{Qian, Gao, and Jagadish}{Qian
  et~al\mbox{.}}{2015}]%
        {qian2015learning}
\bibfield{author}{\bibinfo{person}{Li Qian}, \bibinfo{person}{Jinyang Gao},
  {and} \bibinfo{person}{HV Jagadish}.} \bibinfo{year}{2015}\natexlab{}.
\newblock \showarticletitle{Learning user preferences by adaptive pairwise
  comparison}.
\newblock \bibinfo{journal}{\emph{Proceedings of the VLDB Endowment}}
  \bibinfo{volume}{8}, \bibinfo{number}{11} (\bibinfo{year}{2015}),
  \bibinfo{pages}{1322--1333}.
\newblock


\bibitem[\protect\citeauthoryear{Qin, Luo, Tang, and Li}{Qin
  et~al\mbox{.}}{2020}]%
        {qin2020making}
\bibfield{author}{\bibinfo{person}{Xuedi Qin}, \bibinfo{person}{Yuyu Luo},
  \bibinfo{person}{Nan Tang}, {and} \bibinfo{person}{Guoliang Li}.}
  \bibinfo{year}{2020}\natexlab{}.
\newblock \showarticletitle{Making data visualization more efficient and
  effective: a survey}.
\newblock \bibinfo{journal}{\emph{The VLDB Journal}} \bibinfo{volume}{29},
  \bibinfo{number}{1} (\bibinfo{year}{2020}), \bibinfo{pages}{93--117}.
\newblock


\bibitem[\protect\citeauthoryear{repository}{repository}{2018}]%
        {seedbrepository}
\bibfield{author}{\bibinfo{person}{Rath repository}.}
  \bibinfo{year}{2018}\natexlab{}.
\newblock \bibinfo{howpublished}{\url{https://github.com/snknitin/-SeeDB}}.
\newblock


\bibitem[\protect\citeauthoryear{repository}{repository}{2022}]%
        {rath}
\bibfield{author}{\bibinfo{person}{Rath repository}.}
  \bibinfo{year}{2022}\natexlab{}.
\newblock \bibinfo{howpublished}{\url{https://github.com/Kanaries/Rath}}.
\newblock


\bibitem[\protect\citeauthoryear{Ross}{Ross}{2004}]%
        {ross2004introduction}
\bibfield{author}{\bibinfo{person}{Sheldon~M Ross}.}
  \bibinfo{year}{2004}\natexlab{}.
\newblock \bibinfo{booktitle}{\emph{Introduction to probability and statistics
  for engineers and scientists}}.
\newblock \bibinfo{publisher}{Elsevier}.
\newblock


\bibitem[\protect\citeauthoryear{Roy and Suciu}{Roy and Suciu}{2014}]%
        {RoyS14}
\bibfield{author}{\bibinfo{person}{Sudeepa Roy} {and} \bibinfo{person}{Dan
  Suciu}.} \bibinfo{year}{2014}\natexlab{}.
\newblock \showarticletitle{A formal approach to finding explanations for
  database queries}. In \bibinfo{booktitle}{\emph{SIGMOD}},
  \bibfield{editor}{\bibinfo{person}{Curtis~E. Dyreson},
  \bibinfo{person}{Feifei Li}, {and} \bibinfo{person}{M.~Tamer {\"{O}}zsu}}
  (Eds.). \bibinfo{pages}{1579--1590}.
\newblock


\bibitem[\protect\citeauthoryear{Sarawagi}{Sarawagi}{2001}]%
        {sarawagi2001user}
\bibfield{author}{\bibinfo{person}{Sunita Sarawagi}.}
  \bibinfo{year}{2001}\natexlab{}.
\newblock \showarticletitle{User-cognizant multidimensional analysis}.
\newblock \bibinfo{journal}{\emph{The VLDB Journal}} \bibinfo{volume}{10},
  \bibinfo{number}{2} (\bibinfo{year}{2001}), \bibinfo{pages}{224--239}.
\newblock


\bibitem[\protect\citeauthoryear{Sarawagi, Agrawal, and Megiddo}{Sarawagi
  et~al\mbox{.}}{1998}]%
        {sarawagi1998discovery}
\bibfield{author}{\bibinfo{person}{Sunita Sarawagi}, \bibinfo{person}{Rakesh
  Agrawal}, {and} \bibinfo{person}{Nimrod Megiddo}.}
  \bibinfo{year}{1998}\natexlab{}.
\newblock \showarticletitle{Discovery-driven exploration of OLAP data cubes}.
  In \bibinfo{booktitle}{\emph{EDBT}}.
\newblock


\bibitem[\protect\citeauthoryear{Sch{\"u}tze, Manning, and
  Raghavan}{Sch{\"u}tze et~al\mbox{.}}{2008}]%
        {schutze2008introduction}
\bibfield{author}{\bibinfo{person}{Hinrich Sch{\"u}tze},
  \bibinfo{person}{Christopher~D Manning}, {and} \bibinfo{person}{Prabhakar
  Raghavan}.} \bibinfo{year}{2008}\natexlab{}.
\newblock \bibinfo{booktitle}{\emph{Introduction to information retrieval}}.
  Vol.~\bibinfo{volume}{39}.
\newblock \bibinfo{publisher}{Cambridge University Press Cambridge}.
\newblock


\bibitem[\protect\citeauthoryear{Seleznova, Omidvar-Tehrani, Amer-Yahia, and
  Simon}{Seleznova et~al\mbox{.}}{2020}]%
        {seleznova2020guided}
\bibfield{author}{\bibinfo{person}{Mariia Seleznova}, \bibinfo{person}{Behrooz
  Omidvar-Tehrani}, \bibinfo{person}{Sihem Amer-Yahia}, {and}
  \bibinfo{person}{Eric Simon}.} \bibinfo{year}{2020}\natexlab{}.
\newblock \showarticletitle{Guided exploration of user groups}.
\newblock \bibinfo{journal}{\emph{Proceedings of the VLDB Endowment (PVLDB)}}
  \bibinfo{volume}{13}, \bibinfo{number}{9} (\bibinfo{year}{2020}),
  \bibinfo{pages}{1469--1482}.
\newblock


\bibitem[\protect\citeauthoryear{Sellam and Kersten}{Sellam and
  Kersten}{2016}]%
        {sellam2016cluster}
\bibfield{author}{\bibinfo{person}{Thibault Sellam} {and}
  \bibinfo{person}{Martin Kersten}.} \bibinfo{year}{2016}\natexlab{}.
\newblock \showarticletitle{Cluster-driven navigation of the query space}.
\newblock \bibinfo{journal}{\emph{IEEE Transactions on Knowledge and Data
  Engineering}} \bibinfo{volume}{28}, \bibinfo{number}{5}
  (\bibinfo{year}{2016}), \bibinfo{pages}{1118--1131}.
\newblock


\bibitem[\protect\citeauthoryear{Shafieinejad, Kerschbaum, and
  Ilyas}{Shafieinejad et~al\mbox{.}}{2021}]%
        {shafieinejad2021pcor}
\bibfield{author}{\bibinfo{person}{Masoumeh Shafieinejad},
  \bibinfo{person}{Florian Kerschbaum}, {and} \bibinfo{person}{Ihab~F Ilyas}.}
  \bibinfo{year}{2021}\natexlab{}.
\newblock \showarticletitle{PCOR: Private Contextual Outlier Release via
  Differentially Private Search}. In \bibinfo{booktitle}{\emph{Proceedings of
  the 2021 International Conference on Management of Data}}.
  \bibinfo{pages}{1571--1583}.
\newblock


\bibitem[\protect\citeauthoryear{Shi, Xu, Sun, Shi, and Cao}{Shi
  et~al\mbox{.}}{2020}]%
        {shi2020calliope}
\bibfield{author}{\bibinfo{person}{Danqing Shi}, \bibinfo{person}{Xinyue Xu},
  \bibinfo{person}{Fuling Sun}, \bibinfo{person}{Yang Shi}, {and}
  \bibinfo{person}{Nan Cao}.} \bibinfo{year}{2020}\natexlab{}.
\newblock \showarticletitle{Calliope: Automatic visual data story generation
  from a spreadsheet}.
\newblock \bibinfo{journal}{\emph{IEEE Transactions on Visualization and
  Computer Graphics}} \bibinfo{volume}{27}, \bibinfo{number}{2}
  (\bibinfo{year}{2020}), \bibinfo{pages}{453--463}.
\newblock


\bibitem[\protect\citeauthoryear{Singh, Cafarella, and Jagadish}{Singh
  et~al\mbox{.}}{2016}]%
        {singhdbexplorer}
\bibfield{author}{\bibinfo{person}{Manish Singh}, \bibinfo{person}{Michael~J
  Cafarella}, {and} \bibinfo{person}{HV Jagadish}.}
  \bibinfo{year}{2016}\natexlab{}.
\newblock \showarticletitle{DBExplorer: Exploratory Search in Databases}.
\newblock \bibinfo{journal}{\emph{EDBT}} (\bibinfo{year}{2016}).
\newblock


\bibitem[\protect\citeauthoryear{Srinivasan, Drucker, Endert, and
  Stasko}{Srinivasan et~al\mbox{.}}{2018}]%
        {srinivasan2018augmenting}
\bibfield{author}{\bibinfo{person}{Arjun Srinivasan}, \bibinfo{person}{Steven~M
  Drucker}, \bibinfo{person}{Alex Endert}, {and} \bibinfo{person}{John
  Stasko}.} \bibinfo{year}{2018}\natexlab{}.
\newblock \showarticletitle{Augmenting visualizations with interactive data
  facts to facilitate interpretation and communication}.
\newblock \bibinfo{journal}{\emph{IEEE transactions on visualization and
  computer graphics}} \bibinfo{volume}{25}, \bibinfo{number}{1}
  (\bibinfo{year}{2018}), \bibinfo{pages}{672--681}.
\newblock


\bibitem[\protect\citeauthoryear{\sys\ Repository}{\sys\ Repository}{2022}]%
        {repo}
\bibfield{author}{\bibinfo{person}{\sys\ Repository}.}
  \bibinfo{year}{2022}\natexlab{}.
\newblock \bibinfo{howpublished}{\url{https://github.com/TAU-DB/FEDEx}}.
\newblock


\bibitem[\protect\citeauthoryear{Tang, Han, Yiu, Ding, and Zhang}{Tang
  et~al\mbox{.}}{2017}]%
        {tang2017extracting}
\bibfield{author}{\bibinfo{person}{Bo Tang}, \bibinfo{person}{Shi Han},
  \bibinfo{person}{Man~Lung Yiu}, \bibinfo{person}{Rui Ding}, {and}
  \bibinfo{person}{Dongmei Zhang}.} \bibinfo{year}{2017}\natexlab{}.
\newblock \showarticletitle{Extracting top-k insights from multi-dimensional
  data}. In \bibinfo{booktitle}{\emph{Proceedings of the 2017 ACM International
  Conference on Management of Data}}. \bibinfo{pages}{1509--1524}.
\newblock


\bibitem[\protect\citeauthoryear{ten Cate, Civili, Sherkhonov, and Tan}{ten
  Cate et~al\mbox{.}}{2015}]%
        {ten2015high}
\bibfield{author}{\bibinfo{person}{Balder ten Cate}, \bibinfo{person}{Cristina
  Civili}, \bibinfo{person}{Evgeny Sherkhonov}, {and}
  \bibinfo{person}{Wang-Chiew Tan}.} \bibinfo{year}{2015}\natexlab{}.
\newblock \showarticletitle{High-level why-not explanations using ontologies}.
  In \bibinfo{booktitle}{\emph{Proceedings of the 34th ACM SIGMOD-SIGACT-SIGAI
  Symposium on Principles of Database Systems}}. \bibinfo{pages}{31--43}.
\newblock


\bibitem[\protect\citeauthoryear{Thirumuruganathan, Das, Desai, Amer-Yahia,
  Das, and Yu}{Thirumuruganathan et~al\mbox{.}}{2012}]%
        {thirumuruganathan2012maprat}
\bibfield{author}{\bibinfo{person}{Saravanan Thirumuruganathan},
  \bibinfo{person}{Mahashweta Das}, \bibinfo{person}{Shrikant Desai},
  \bibinfo{person}{Sihem Amer-Yahia}, \bibinfo{person}{Gautam Das}, {and}
  \bibinfo{person}{Cong Yu}.} \bibinfo{year}{2012}\natexlab{}.
\newblock \showarticletitle{Maprat: Meaningful explanation, interactive
  exploration and geo-visualization of collaborative ratings}.
\newblock \bibinfo{journal}{\emph{Proceedings of the VLDB Endowment (PVLDB)}}
  \bibinfo{volume}{5}, \bibinfo{number}{12} (\bibinfo{year}{2012}),
  \bibinfo{pages}{1986--1989}.
\newblock


\bibitem[\protect\citeauthoryear{van Leeuwen}{van Leeuwen}{2010}]%
        {van2010maximal}
\bibfield{author}{\bibinfo{person}{Matthijs van Leeuwen}.}
  \bibinfo{year}{2010}\natexlab{}.
\newblock \showarticletitle{Maximal exceptions with minimal descriptions}.
\newblock \bibinfo{journal}{\emph{Data Mining and Knowledge Discovery}}
  \bibinfo{volume}{21}, \bibinfo{number}{2} (\bibinfo{year}{2010}),
  \bibinfo{pages}{259--276}.
\newblock


\bibitem[\protect\citeauthoryear{Vartak, Rahman, Madden, Parameswaran, and
  Polyzotis}{Vartak et~al\mbox{.}}{2015}]%
        {VartakRMPP15}
\bibfield{author}{\bibinfo{person}{Manasi Vartak}, \bibinfo{person}{Sajjadur
  Rahman}, \bibinfo{person}{Samuel Madden}, \bibinfo{person}{Aditya~G.
  Parameswaran}, {and} \bibinfo{person}{Neoklis Polyzotis}.}
  \bibinfo{year}{2015}\natexlab{}.
\newblock \showarticletitle{{SEEDB:} Efficient Data-Driven Visualization
  Recommendations to Support Visual Analytics}.
\newblock \bibinfo{journal}{\emph{Proc. {VLDB} Endow.}} \bibinfo{volume}{8},
  \bibinfo{number}{13} (\bibinfo{year}{2015}), \bibinfo{pages}{2182--2193}.
\newblock


\bibitem[\protect\citeauthoryear{Wang, Sun, Zhang, Cui, Xu, Ma, and Zhang}{Wang
  et~al\mbox{.}}{2019}]%
        {wang2019datashot}
\bibfield{author}{\bibinfo{person}{Yun Wang}, \bibinfo{person}{Zhida Sun},
  \bibinfo{person}{Haidong Zhang}, \bibinfo{person}{Weiwei Cui},
  \bibinfo{person}{Ke Xu}, \bibinfo{person}{Xiaojuan Ma}, {and}
  \bibinfo{person}{Dongmei Zhang}.} \bibinfo{year}{2019}\natexlab{}.
\newblock \showarticletitle{Datashot: Automatic generation of fact sheets from
  tabular data}.
\newblock \bibinfo{journal}{\emph{IEEE transactions on visualization and
  computer graphics}} \bibinfo{volume}{26}, \bibinfo{number}{1}
  (\bibinfo{year}{2019}), \bibinfo{pages}{895--905}.
\newblock


\bibitem[\protect\citeauthoryear{Wongsuphasawat, Moritz, Anand, Mackinlay,
  Howe, and Heer}{Wongsuphasawat et~al\mbox{.}}{2016}]%
        {wongsuphasawat2016voyager}
\bibfield{author}{\bibinfo{person}{Kanit Wongsuphasawat},
  \bibinfo{person}{Dominik Moritz}, \bibinfo{person}{Anushka Anand},
  \bibinfo{person}{Jock Mackinlay}, \bibinfo{person}{Bill Howe}, {and}
  \bibinfo{person}{Jeffrey Heer}.} \bibinfo{year}{2016}\natexlab{}.
\newblock \showarticletitle{Voyager: Exploratory analysis via faceted browsing
  of visualization recommendations}.
\newblock \bibinfo{journal}{\emph{TVCG}} (\bibinfo{year}{2016}).
\newblock


\bibitem[\protect\citeauthoryear{Wu and Madden}{Wu and Madden}{2013}]%
        {0002M13}
\bibfield{author}{\bibinfo{person}{Eugene Wu} {and} \bibinfo{person}{Samuel
  Madden}.} \bibinfo{year}{2013}\natexlab{}.
\newblock \showarticletitle{Scorpion: Explaining Away Outliers in Aggregate
  Queries}.
\newblock \bibinfo{journal}{\emph{Proc. {VLDB} Endow.}} \bibinfo{volume}{6},
  \bibinfo{number}{8} (\bibinfo{year}{2013}), \bibinfo{pages}{553--564}.
\newblock


\bibitem[\protect\citeauthoryear{Yan and He}{Yan and He}{2020}]%
        {YanH20}
\bibfield{author}{\bibinfo{person}{Cong Yan} {and} \bibinfo{person}{Yeye He}.}
  \bibinfo{year}{2020}\natexlab{}.
\newblock \showarticletitle{Auto-Suggest: Learning-to-Recommend Data
  Preparation Steps Using Data Science Notebooks}. In
  \bibinfo{booktitle}{\emph{SIGMOD}}. \bibinfo{pages}{1539--1554}.
\newblock


\bibitem[\protect\citeauthoryear{Zegarra, Ipenza, Omidvar-Tehrani, Moreira,
  Amer-Yahia, and Comba}{Zegarra et~al\mbox{.}}{2020}]%
        {zegarra2020visual}
\bibfield{author}{\bibinfo{person}{Fabian~Colque Zegarra},
  \bibinfo{person}{Juan C~Carbajal Ipenza}, \bibinfo{person}{Behrooz
  Omidvar-Tehrani}, \bibinfo{person}{Viviane~P Moreira}, \bibinfo{person}{Sihem
  Amer-Yahia}, {and} \bibinfo{person}{Jo{\~a}o~LD Comba}.}
  \bibinfo{year}{2020}\natexlab{}.
\newblock \showarticletitle{Visual exploration of rating datasets and user
  groups}.
\newblock \bibinfo{journal}{\emph{Future Generation Computer Systems}}
  \bibinfo{volume}{105} (\bibinfo{year}{2020}), \bibinfo{pages}{547--561}.
\newblock


\end{thebibliography}
